\documentclass{emulateapj}
\usepackage{natbib,apjfonts}
\newcommand{\HI}{\ion{H}{1}}
\newcommand{\HII}{\ion{H}{2}}

\newcommand{\kms}{\mbox{km~s$^{-1}$}}

\newcommand{\siggas}{\mbox{$\Sigma_g$}}
\newcommand{\sigsfr}{\mbox{$\Sigma_{\rm SFR}$}}

\shorttitle{Timescale for Star Formation in Galaxies}
\submitted{Received 6 Aug 2009; Accepted 8 Sep 2009}
\journalinfo{To Appear in the Astrophysical Journal}

\begin{document}

\title{On the Timescale for Star Formation in Galaxies}

\author{Tony Wong}
\affil{Astronomy Department, University of Illinois, Urbana, IL 61801}
\email{wongt@astro.uiuc.edu}

\begin{abstract}

The timescale for star formation, a measure of how quickly neutral gas is being converted to stars, is considerably longer than typical dynamical timescales associated with a galactic disk.  For purposes of modeling galaxy evolution, however, it would be extremely attractive if the star formation timescale were proportional to an easily derived dynamical timescale.   We compare estimates of the star formation timescale within nearby galaxies, based on the work of \citet{Leroy:08} and existing BIMA SONG CO data, with three simple forms of the dynamical time: the orbital time, the free-fall time at the midplane density, and the disk Jeans time (the growth time for gravitational instabilities in a disk).  When taking into account the gravity of the stellar disk in an approximate way, all three timescales show correlations with the star formation timescale, though none of the correlations can be accurately described as linear.  Systematic errors in estimating appropriate gas masses and the stellar velocity dispersion may obscure an underlying correlation, but we focus instead on a model where the timescale for H$_2$ formation from \HI\ is decoupled from the timescale for star formation from H$_2$.  The Jeans time correlates well with the first of these timescales, but the relationship is still non-linear, and requires a characteristic GMC lifetime that increases toward galaxy centers.

\end{abstract}


\section{Introduction}

In recent years increasing attention has been focused on the question of how the star formation rate (SFR) in disk galaxies is determined, as this remains one of the key parameters difficult to specify {\it ab initio} in cosmological simulations.  By far the most common prescription is the ``Schmidt law'', which posits that the SFR per unit area (or volume) scales with the gas surface (or volume) density raised to a fixed power, e.g.
\begin{equation}
\sigsfr = k (\Sigma_g)^N \;,
\label{eqn:kslaw}
\end{equation}
as suggested by \citet{Schmidt:59} and first demonstrated empirically for external galaxies by \citet{KC:89}.  (Hereafter we use $N$ to refer to the power law relating surface densities and $n$ for that relating volume densities.)  Since the surface density $\Sigma$ is simply the projection of volume density $\rho$ through the disk, given a constant scale height for gas and star formation one should obtain the same power law indices for surface and volume densities.  This has led to the suggestion that the observed index of $N$$\approx$1.5 \citep{KC:98a} can be understood if the SFR per unit gas mass, $\dot{\rho}_*/\rho_g$ (hereafter the star formation efficiency or SFE), is proportional to the free-fall rate, $\omega_{\rm ff} \propto (G\rho_g)^{1/2}$.  As is well known \citep[e.g.,][]{Zuckerman:74}, the actual proportionality constant must be much smaller than 1 in order to account for the observed inefficiency of star formation (i.e., the relatively long timescale for consuming all interstellar gas compared to the free-fall time).

On the other hand, the assumption of a constant gas scale height is unlikely to be correct; the \HI\ disks of galaxies are observed to flare outwards at large  radius \citep[e.g.,][]{Olling:96}, while the vertical gas velocity dispersion $c_g$ appears to be fairly constant \citep{Shostak:84}.  For a purely gaseous disk in gravitational equilibrium, the scale height can be approximated as $h_g=c_g^2/(\pi G\Sigma_g)$ \citep{vdK:81a}, and the vertical crossing time $h_g/c_g$=$c_g/(\pi G\Sigma_g)$ is analogous to the free-fall time.  If the SFE indeed scales with the free-fall rate, one therefore expects a power-law index $N$=2 for surface density.

The widely adopted $N$=1.4 Schmidt law of \citet{KC:98a} is based on global disk averages augmented with spatially resolved images of infrared-luminous starbursts.  In general, observational tests of the Schmidt law {\it within} galaxies have been limited by the scarcity of spatially resolved CO and \HI\ maps of nearby galaxies.  \citet{Wong:02} investigated the azimuthally averaged star formation law in a sample of seven CO-luminous galaxies from the BIMA Survey of Nearby Galaxies (BIMA SONG).  Using their favored model for correcting H$\alpha$ emission for extinction, they estimated $N \approx 1.75 \pm 0.25$ when comparing the surface densities of total gas and SFR.  \citet{Boissier:03} employed somewhat different assumptions for inferring gas and SFR surface densities, but arrived at a similar result of $N \approx 2$ for a sample of 16 galaxies.  On the other hand, \citet{Heyer:04} found $N \approx 3.3$ in M33, a consequence of the relatively flat radial gas profile in this \HI-dominated galaxy.  More recently, \citet{Bigiel:08} and \citet{Leroy:08} have exploited improvements in \HI, CO, and infrared imaging to examine the Schmidt law across a wider range of galactocentric radii and in both CO-luminous and CO-faint galaxies.  \citet{Bigiel:08} find an average index of $N \approx 1.85 \pm 0.70$ for a sample of seven galaxies, with a factor of $\sim$2 scatter about the mean relation.  \citet{Leroy:08} identify two regimes, an H$_2$-dominated regime (in the inner disks of massive spirals) where the SFE is roughly constant, and an \HI-dominated regime (in outer disks and dwarfs) where the SFE shows a strong positive correlation with the {\it stellar} surface density, $\Sigma_*$.

The high degree of variability in the observed power law index $N$, and the lack of any clear correlation between the SFR and \HI\ surface density in spatially resolved studies (see also \citealt{Kennicutt:07}), have led many authors to conclude that the Schmidt law is largely a property of the molecular gas, since restricting the comparison to CO fluxes yields a narrower range of values for $N$ ranging from 1--1.4 \citep{Wong:02,Heyer:04,Bigiel:08}.  Indeed the densest molecular gas, traced by HCN $J$=1--0 emission, appears to be converted into stars on a roughly constant timescale of $\sim$$10^8$ yr \citep{Gao:04b,Wu:05}.  \citet{Krumholz:07} argue that this timescale is proportional to the free-fall time at the critical density of the HCN line, which is higher than the median density of the ISM: thus all HCN-emitting regions tend to have this density and free-fall time.  However, focusing solely on high-density gas leaves unanswered the question of how the H$_2$ or ``dense'' gas content is determined, and whether galactic-scale processes (involving one or more dynamical timescales) are involved.  Furthermore, if the gas traced by CO emission is part of a continuous range in density, one might reasonably expect that the star formation rate can be predicted from the average gas density, including the atomic gas.

Our focus in this paper is on obtaining simple estimates of three dynamical timescales for a galactic disk composed of gas and stars, and comparing these estimates with recent data.  Although none of these timescales appear to yield a satisfactory prediction for the SFR, the inclusion of the stellar component appears to be important in reproducing some of the observed trends.  Our estimates are based on azimuthal averages, and thus may fail to reflect the appropriate {\it local} dynamical timescales.  Moreover, the difficulty of measuring stellar and gaseous velocity dispersions, and the frequent presence of a stellar bulge in regions of bright CO emission, will complicate the evaluation of the free-fall time.  Thus, this approach must be tested and refined in nearby systems where star formation and GMCs can be spatially resolved.

\begin{figure*}[t]
\begin{center}
\includegraphics[bb=42 157 540 675,clip,width={.48\textwidth}]{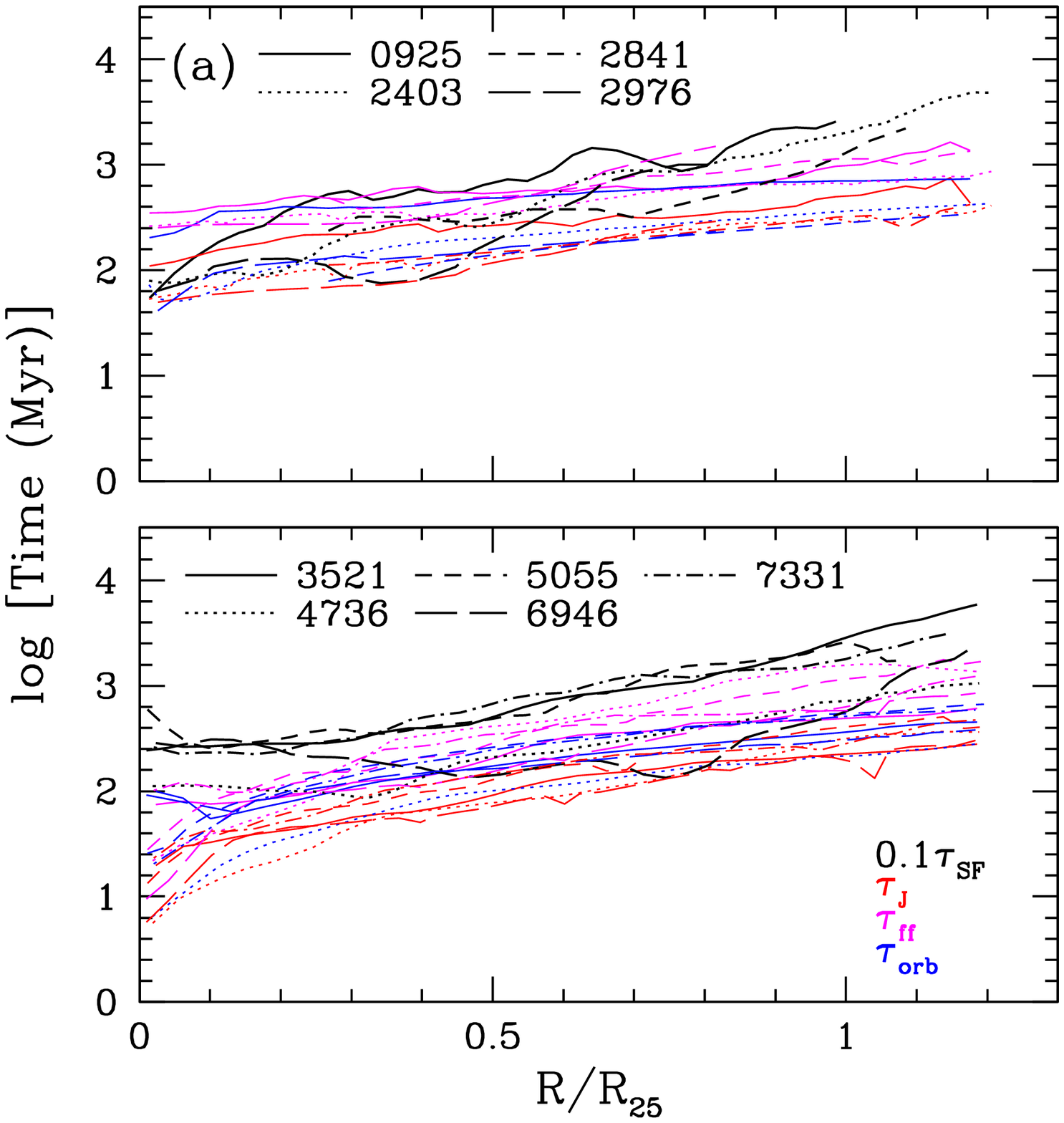}\hfill
\includegraphics[bb=42 157 540 675,clip,width={.48\textwidth}]{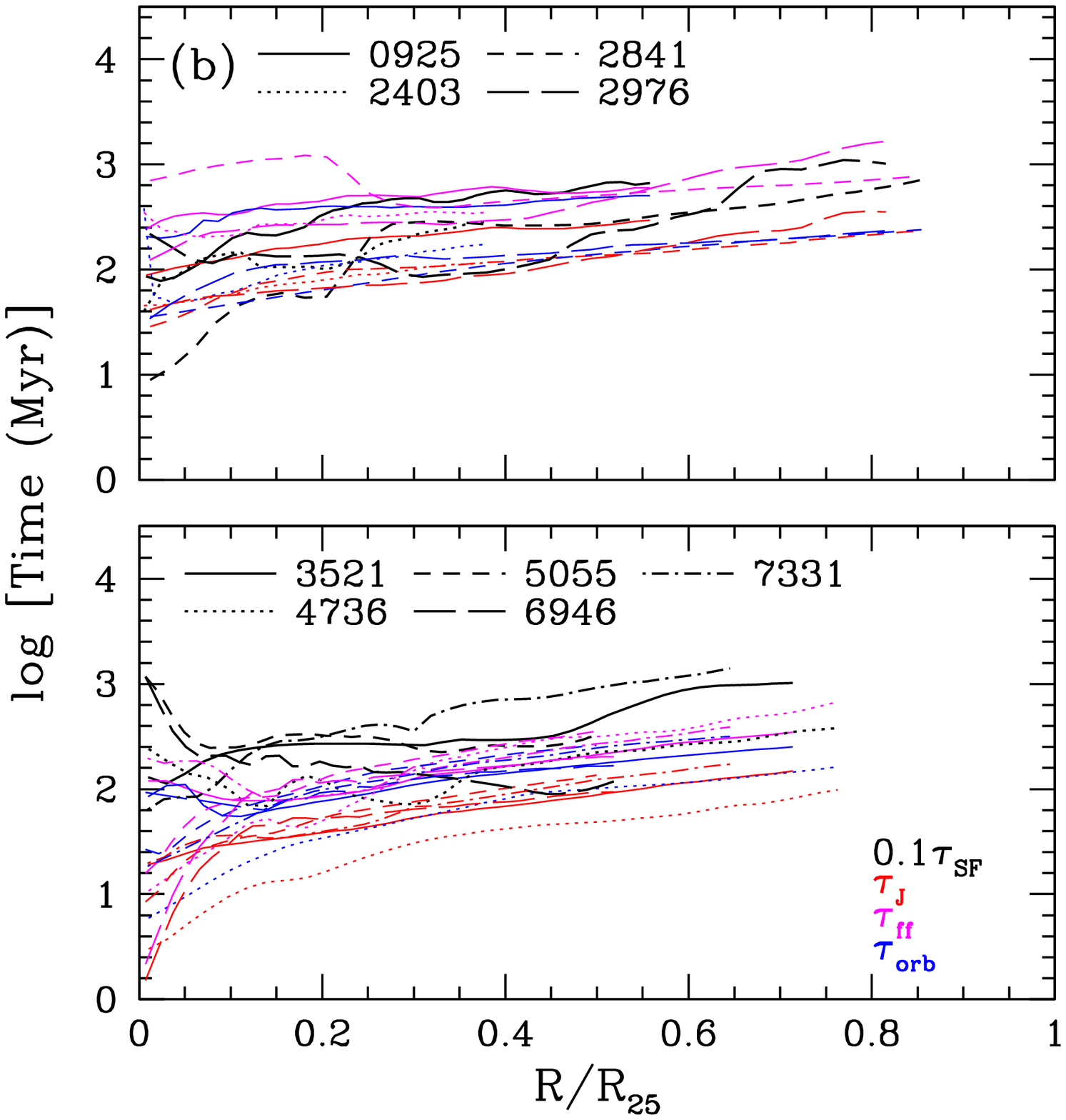}
\end{center}
\caption{Star formation and dynamical timescales plotted vs.\ radius normalized to $R_{25}$.  {\it (Left)} Timescales are based on the radial gas and stellar profiles of L08 and the HI rotation curves of \citet{deBlok:08}.  {\it (Right)} Same, except BIMA CO profiles are used for H$_2$ and 2MASS $K_s$-band profiles for $\Sigma_*$.  Line colors indicate different timescales, line styles indicate different galaxies.}
\label{fig:tradplt}
\end{figure*}

\section{Three Dynamical Rates}

The fundamental dynamical rates for a galactic disk are the orbital rate $\Omega$ and the Jeans rate $\omega_{\rm J} = \pi G\Sigma/c_s$, where $\Sigma$ is the mass surface density and $c_s$ is the sound speed (typically replaced by an ``effective'' sound speed that includes the contributions of turbulent and magnetic pressure).  The two are related via the axisymmetric instability parameter $Q \equiv \kappa c_s/(\pi G\Sigma)$:
\[\omega_{\rm J} = \frac{\kappa}{Q} \propto \frac{\Omega}{Q}\;,\]
where we have assumed $\kappa \propto \Omega$ (e.g., a flat rotation curve).  The Jeans rate is the natural growth rate for gravitational instability in a disk, although it should be noted that axisymmetric (ring) instabilities will be suppressed if $Q>1$, and even for $Q<1$ their growth rate is suppressed by a factor of $\sqrt{1-Q^2}$ \citep{Elm:92b}.  In addition to these two rates, there is a third rate, the free-fall rate $\omega_{\rm ff} \sim (G\rho)^{1/2}$, which is estimated from the local volume density.  For an isothermal disk in hydrostatic equilibrium, the midplane density can be derived from the scale height $h=c_s^2/(\pi G\Sigma)$ using $\rho_0 = \Sigma/(2h)$, so it follows that $\omega_{\rm ff} \sim \omega_{\rm J}$ to within factors of a few.

An approximate adjustment for the case of a two-component disk of gas and stars can be made by identifying the Jeans rate with the vertical oscillation rate in the disk, shown by \citet{Talbot:75} to be:
\begin{eqnarray}
\omega_{\rm J,sg} & = & \pi G \left(\frac{\Sigma_*}{c_*} + \frac{\Sigma_g}{c_g}\right)\\
& = & \frac{\pi G\Sigma_g}{c_g} \left(1+\frac{c_g}{c_*}\frac{\Sigma_*}{\Sigma_g}\right)
\end{eqnarray}
where $c_*$ is the {\it vertical} stellar velocity dispersion.

Similarly, the free-fall rate for gas at the equilibrium midplane density $\rho_0$ can be written as
\begin{equation}
\omega_{\rm ff,sg} = \left(\frac{32\pi}{3}G\rho_{\rm 0,sg}\right)^{1/2} = \frac{4\pi G \Sigma_g}{\sqrt{3}c_g} \left(1+\frac{c_g}{c_*}\frac{\Sigma_*}{\Sigma_g}\right)^{1/2}
\end{equation}
using the approximation for the midplane gas density derived by \citet{Leroy:08} from work by \citet{Elmegreen:89} and \citet{Krumholz:05}:
\begin{equation}
\rho_{\rm 0,sg} = \frac{\pi G\Sigma_g^2}{2c_g^2}\left(1 + \frac{c_g}{c_*}\frac{\Sigma_*}{\Sigma_g}\right) = \eta\rho_0 \;,
\end{equation}
where $\eta = 1+(c_g/c_*)(\Sigma_*/\Sigma_g)$ is a correction factor that represents the additional compression of the gas layer due to the gravity of the stellar disk.  

The importance of the correction factor $\eta$ lies in its possible effect on the observed star formation law.  For a star formation rate $\propto \Sigma_g \omega_{\rm J}$ or $\propto \Sigma_g \omega_{\rm ff}$, one expects $\sigsfr \propto \Sigma_g^2$ for constant gas velocity dispersion $c_g$.  However, including the effect of the stellar disk increases the star formation rate by a factor of $\eta$ (adopting the Jeans rate) or $\eta^{1/2}$ (adopting the free-fall rate).  Since $\eta$ generally anti-correlates with $\Sigma_g$, the tendency is to reduce the Schmidt power law below $N$=2, by an amount which depends on the relative dominance of the stellar disk (see \S\ref{sec:schmidt} for further discussion).

While the orbital rate is not directly tied to growth rates for gravitational instabilities, it may be more straightforward to estimate in numerical simulations since it reflects the gravitational potential of the galaxy as a whole.  In addition, if star formation feedback constrains the $Q$ parameter towards a narrow range of values, whereas the assumption of constant $c_g$ is {\it not} generally valid, then a star formation law based on the orbital rate may turn out to be more useful in practice, since $\Omega/Q$ may be easier to estimate than $\Sigma_g/c_g$.  On the other hand, if $Q$ varies with radius, the orbital and Jeans rates may be substantially different.  Note that when the stellar disk is taken into account, the ratio $\omega_{\rm J,sg}/\Omega \propto \eta/Q = Q^{-1} + Q_*^{-1} \equiv Q_{\rm eff}^{-1}$, which is approximately the instability parameter for a combined disk of gas and stars derived by \citet{Wang:94} (ignoring the anisotropy of the stellar velocity dispersion).  Thus, a proportionality between $\omega_{\rm J,sg}$ and $\Omega$ is maintained as long as $Q_{\rm eff}$ is roughly constant.

Hereafter we omit the subscript `sg' so that timescales are calculated including the correction factor for the stellar disk.

\begin{figure*}
\begin{center}
\includegraphics[bb=42 157 540 675,clip,width={.48\textwidth}]{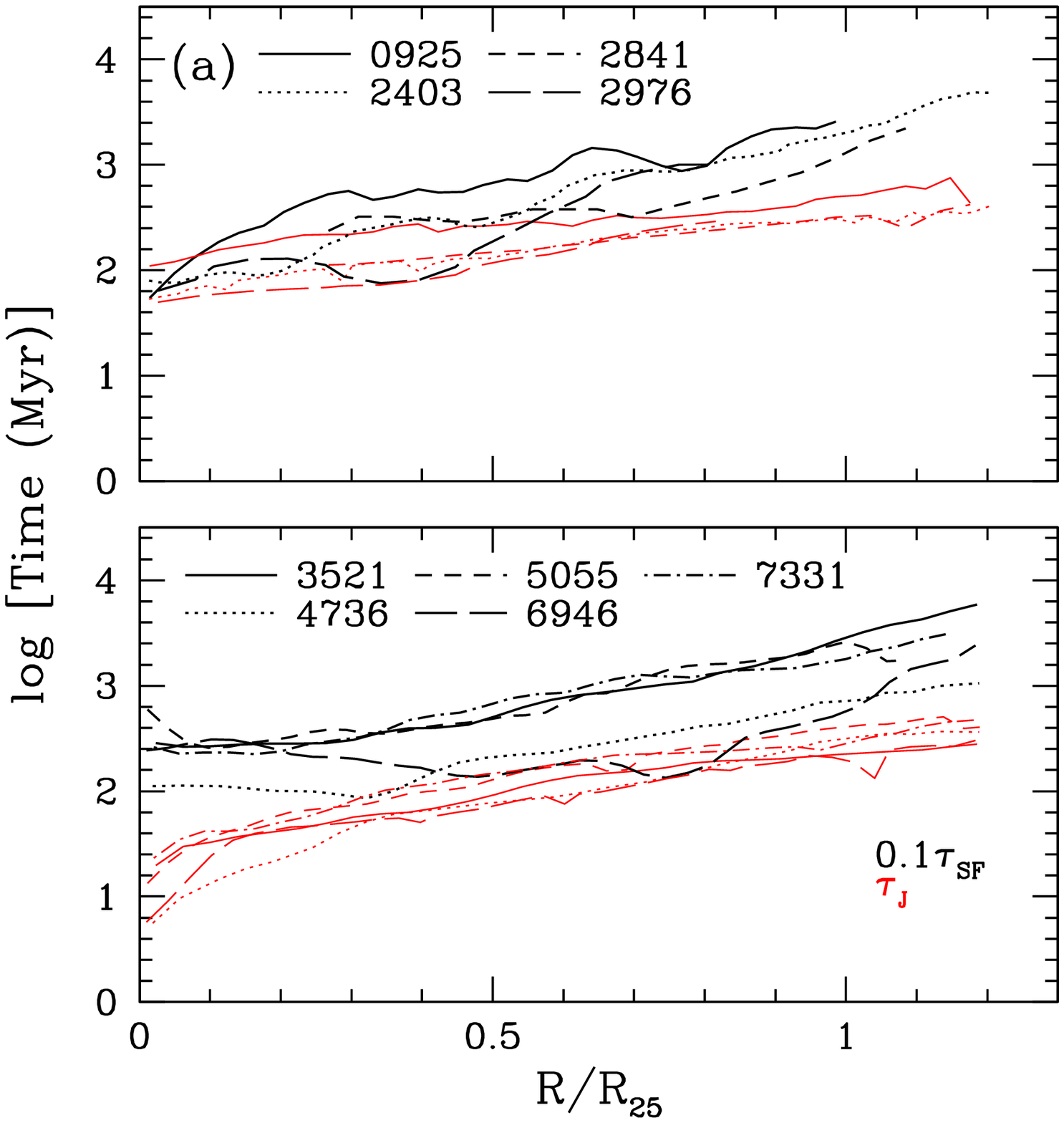}\hfill
\includegraphics[bb=42 157 540 675,clip,width={.48\textwidth}]{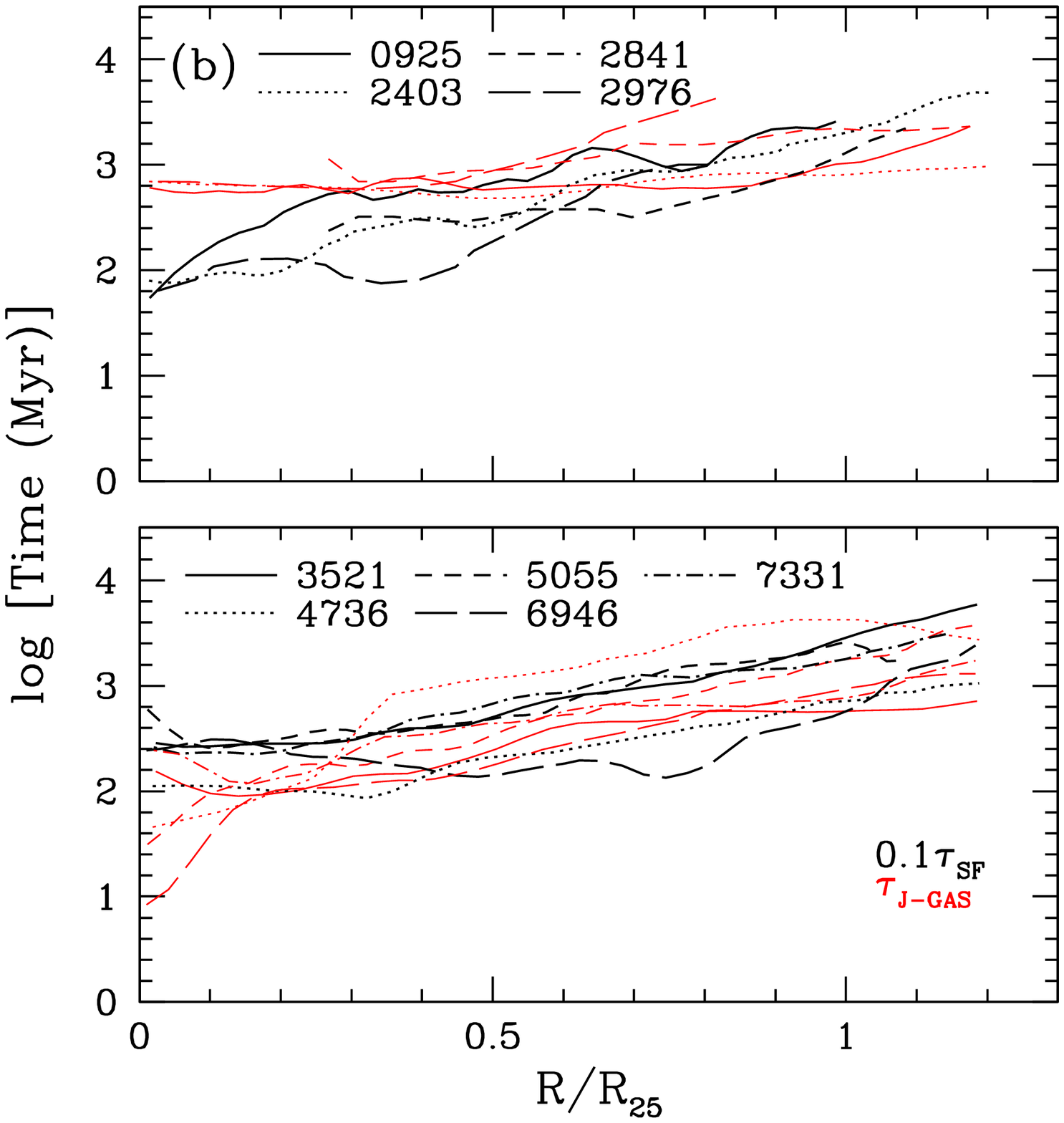}
\end{center}
\caption{Comparison of Jeans timescales (in red) calculated from both stellar and gas profiles ({\it left}) and gas profiles only ({\it right}) from L08.}
\label{fig:jeans}
\end{figure*}

\section{Observational Data}

We primarily make use of radial profiles of $\Sigma_*$, \sigsfr, and $\Sigma_g$ [composed of $\Sigma$(\HI) and $\Sigma$(H$_2$)] published by \citeauthor{Leroy:08} (\citeyear{Leroy:08}, hereafter L08), which we briefly describe here.  
The profiles of $\Sigma_*$ are derived from {\it Spitzer} 3.6 $\mu$m maps from SINGS \citep{Kennicutt:03} using an empirical scaling factor from $K$ to 3.6 $\mu$m and a $K$-band mass-to-light ratio of $\Upsilon_*^K = 0.5\, M_\odot/L_{\odot,K}$.  
\sigsfr\ is derived from a combination of FUV images from the {\it GALEX} Nearby Galaxies Survey \citep{GildePaz:07} and 24 $\mu$m images from SINGS.  $\Sigma$(H$_2$) is derived from CO(2--1) images from the IRAM 30m telescope assuming $I_{\rm CO}$(2--1)/$I_{\rm CO}$(1--0)=0.8 and a Galactic value of the CO-to-H$_2$ conversion factor, $X_{\rm CO}$=$2 \times 10^{20}$ cm$^{-2}$ (K km s$^{-1}$)$^{-1}$.  
$\Sigma$(\HI) is derived from the VLA THINGS survey \citep{Walter:08} assuming the 21-cm line is optically thin.  
We choose three galaxies from L08's ``dwarf'' sample (NGC 925, 2403, and 2976) and six galaxies from their ``spiral'' sample (NGC 2841, 3521, 4736, 5055, 6946, and 7331).  
Radial profiles of the dwarf sample were obtained at resolutions of 9\arcsec--26\arcsec\ (chosen to achieve a spatial resolution of 400 pc) while radial profiles of the spiral sample are at resolutions of 11\arcsec--35\arcsec\ (chosen to achieve a spatial resolution of 800 pc).

We use the BIMA Survey of Nearby Galaxies (BIMA SONG; \citealt{Helfer:03}) as an alternative source for CO data.  
The publicly available CO(1--0) data cubes for the nine sample galaxies were convolved to the resolution of the THINGS \HI\ data (typically $\sim$10\arcsec), or, in cases where the \HI\ data have higher resolution, both CO and \HI\ were convolved to a resolution of 8\arcsec.  
Radial profiles were derived by averaging in elliptical rings using the center positions, inclinations, and position angles of L08.  
\HI\ profiles were derived from the integrated intensity maps provided on the THINGS website.  
CO profiles were derived from integrated intensity maps produced using a smooth-and-mask technique: a mask was generated using a 3$\sigma$ threshold in the cube after smoothing to 15\arcsec, and unmasked pixels exceeding 2$\sigma$ were summed in velocity.  
For galaxies with very weak CO emission (NGC 925, 2403, 2841, and 2976), the mask was generated using a 6$\sigma$ threshold in the \HI\ cube, again smoothed to 15\arcsec.  
We verified that the profiles derived from the masked cubes were consistent (within the uncertainties) with profiles derived by direct summation of the channel maps.  
To derive the total (\HI+H$_2$) gas profile we extrapolated the CO profile beyond the last reliably measured point (typically $R$=90\arcsec) using an exponential fit to the region $R \gtrsim 40\arcsec$.

To obtain a $\Sigma_*$ profile appropriate for the higher resolution of the SONG data, we use $K$-band images from the 2MASS Large Galaxy Atlas \citep{Jarrett:03}, available on the IRSA website, assuming a native resolution of 2\farcs5 and applying a Gaussian smoothing kernel to achieve the same resolution as the CO and \HI\ images.  The radial profile is taken as the median surface brightness in each elliptical annulus, corrected for inclination and converted to a mass surface density using a mass-to-light ratio of $\Upsilon_*^K = 0.5\, M_\odot/L_{\odot,K}$ for consistency with L08.
For \sigsfr\ we simply adopt the profiles provided by L08.  Note that these are at somewhat lower resolution than the CO, \HI, and stellar profiles, and so comparisons of $\tau_{\rm SF}$ with dynamical timescales must be made with caution, particularly near the galaxy centers.

In order to estimate dynamical timescales, we also require radial profiles for $\Omega$, $c_g$, and $c_*$.  We use the THINGS \HI\ rotation curves of \citet{deBlok:08} to derive $\Omega$, and adopt a constant value of 10 \kms\ for $c_g$, similar to that adopted by L08.  
Lacking information about the radial variation of the stellar velocity dispersion, we follow standard practice (e.g., L08) by assuming a constant stellar scale height $h_*$, and an isothermal disk such that
\[h_* = \frac{c_{*}^2}{2\pi G \Sigma_*}\;.\]
Here we define $h_*$ such that an isothermal disk has $\rho(z) \propto {\rm sech}^2(z/2h_*)$, which asymptotically approaches $\exp(-z/h_*)$ at large $z$ \citep{vdK:88}.  We estimate the scale height using the empirical relation between radial and vertical scalelengths of \citet{Kregel:02},
\[\left\langle\frac{l_*}{h_*}\right\rangle = 7.3 \pm 2.2\;,\]
taking for $l_*$ the exponential $K$-band scalelengths given by the 2MASS LGA (parameter $\alpha$).  We then use this assumed value of $h_*$ along with the $\Sigma_*$ profiles to derive the radial variation in $c_*$.  While the assumption of a sech$^{2}$ disk is questionable, with numerical calculations by \citet{Banerjee:07} suggesting that in the presence of a gaseous disk the vertical stellar profile is closer to exponential than sech$^{2}$, adopting different vertical profiles has little effect on the relation between $c_*$ and $\Sigma_*$ for a given $h_*$ \citep{vdK:88}.

\begin{figure*}
\begin{center}
\includegraphics[bb=42 157 540 675,clip,width={.48\textwidth}]{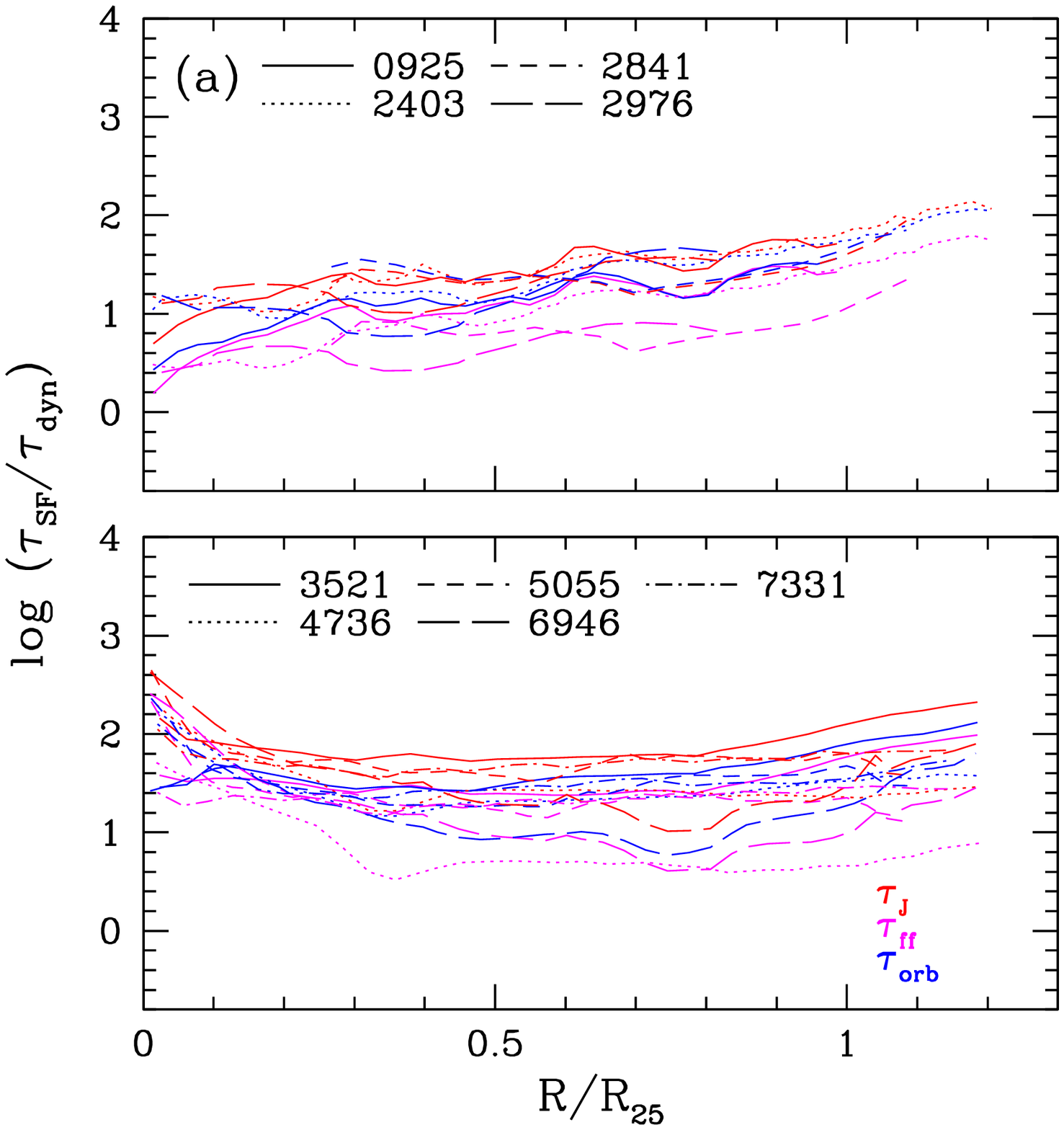}\hfill
\includegraphics[bb=42 157 540 675,clip,width={.48\textwidth}]{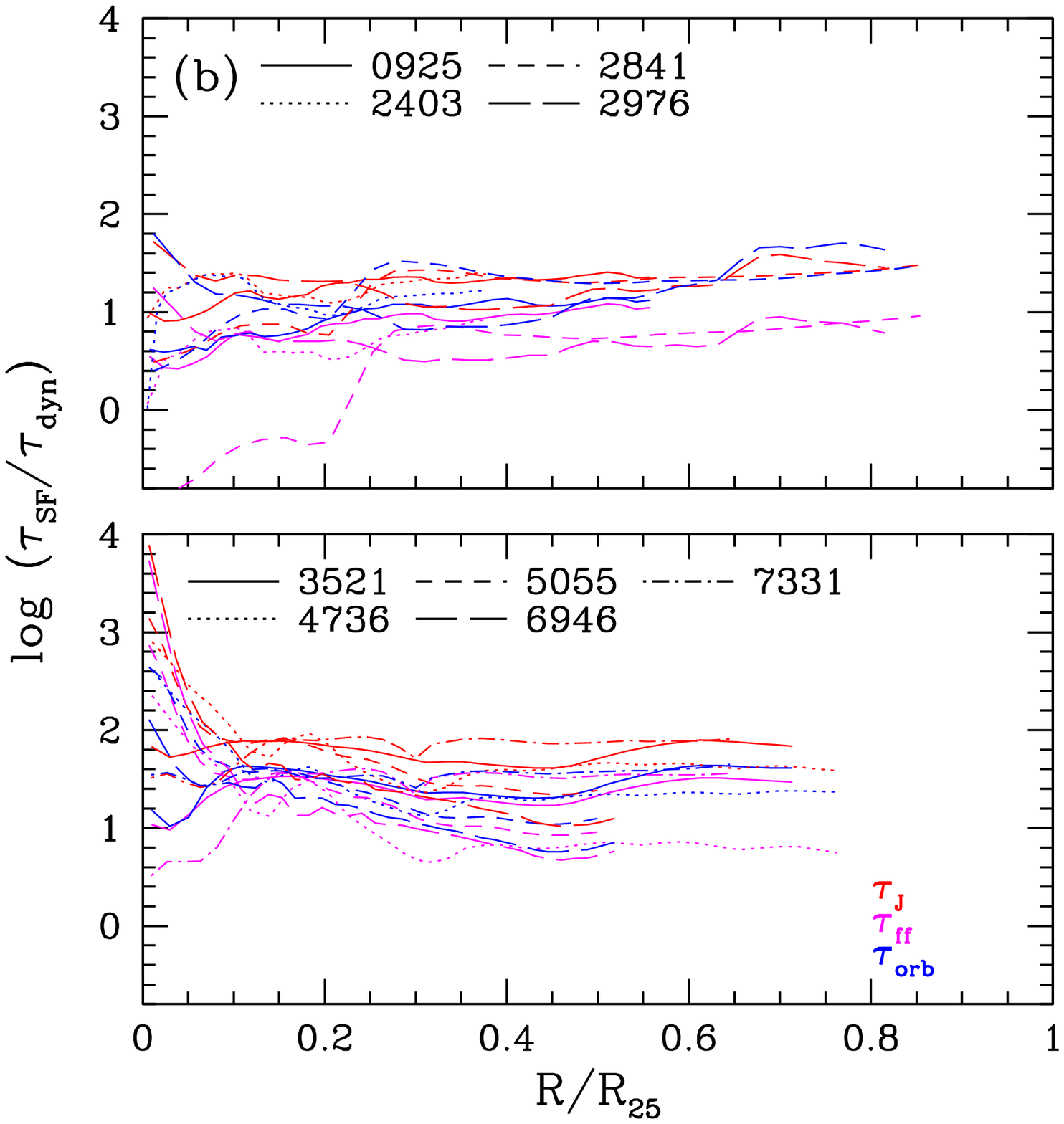}
\end{center}
\caption{Ratios of star formation to various dynamical timescales plotted vs.\ radius normalized to $R_{25}$.  Data are from L08 {\it (left)} and BIMA SONG {\it (right)} as in Fig.~\ref{fig:tradplt}.  Line colors indicate different timescales, line styles indicate different galaxies.}
\label{fig:tnorm}
\end{figure*}

\begin{figure*}[t]
\begin{center}
\includegraphics[bb=36 156 540 665,clip,width={.32\textwidth}]{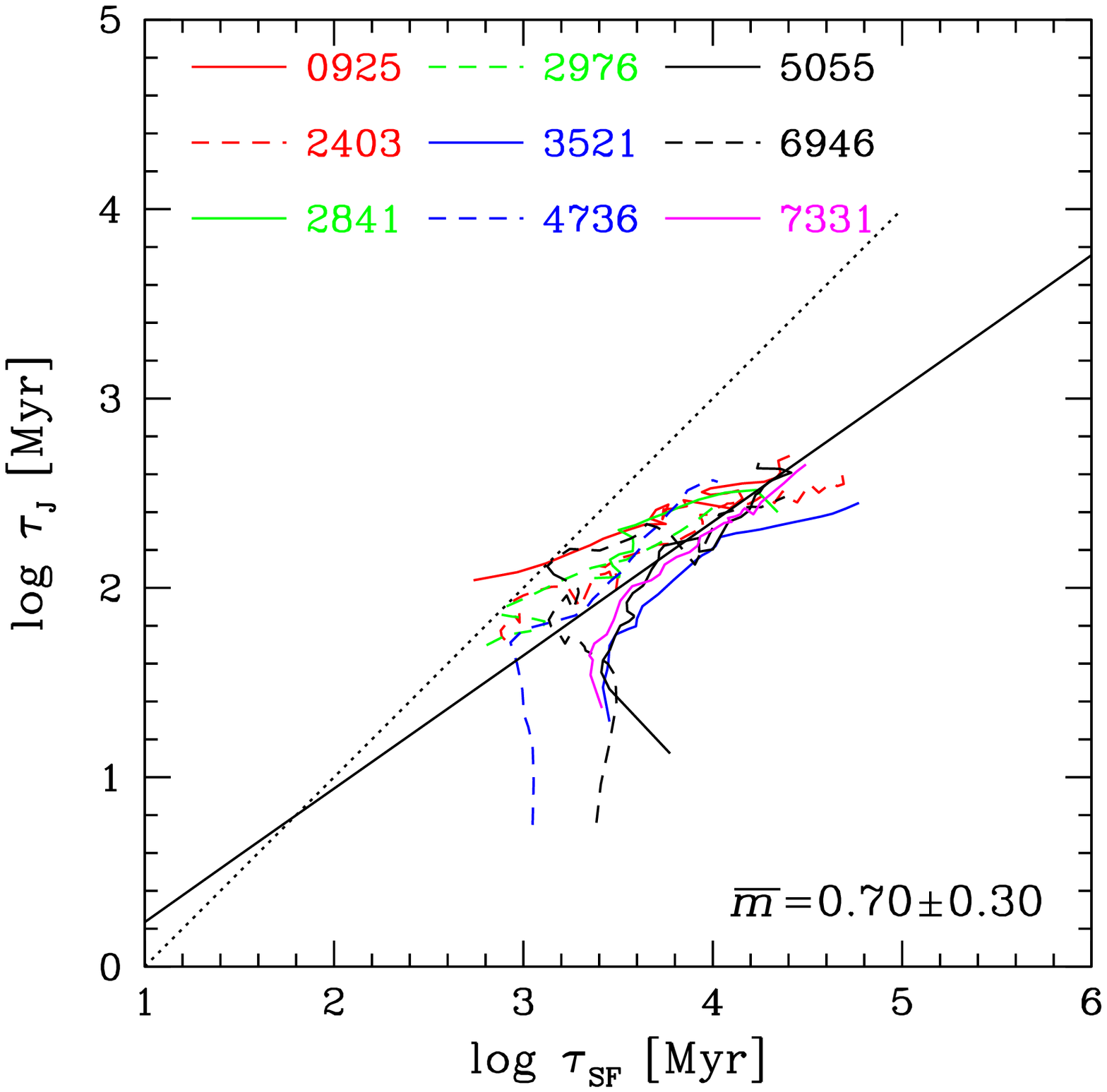}\hfill
\includegraphics[bb=36 156 540 665,clip,width={.32\textwidth}]{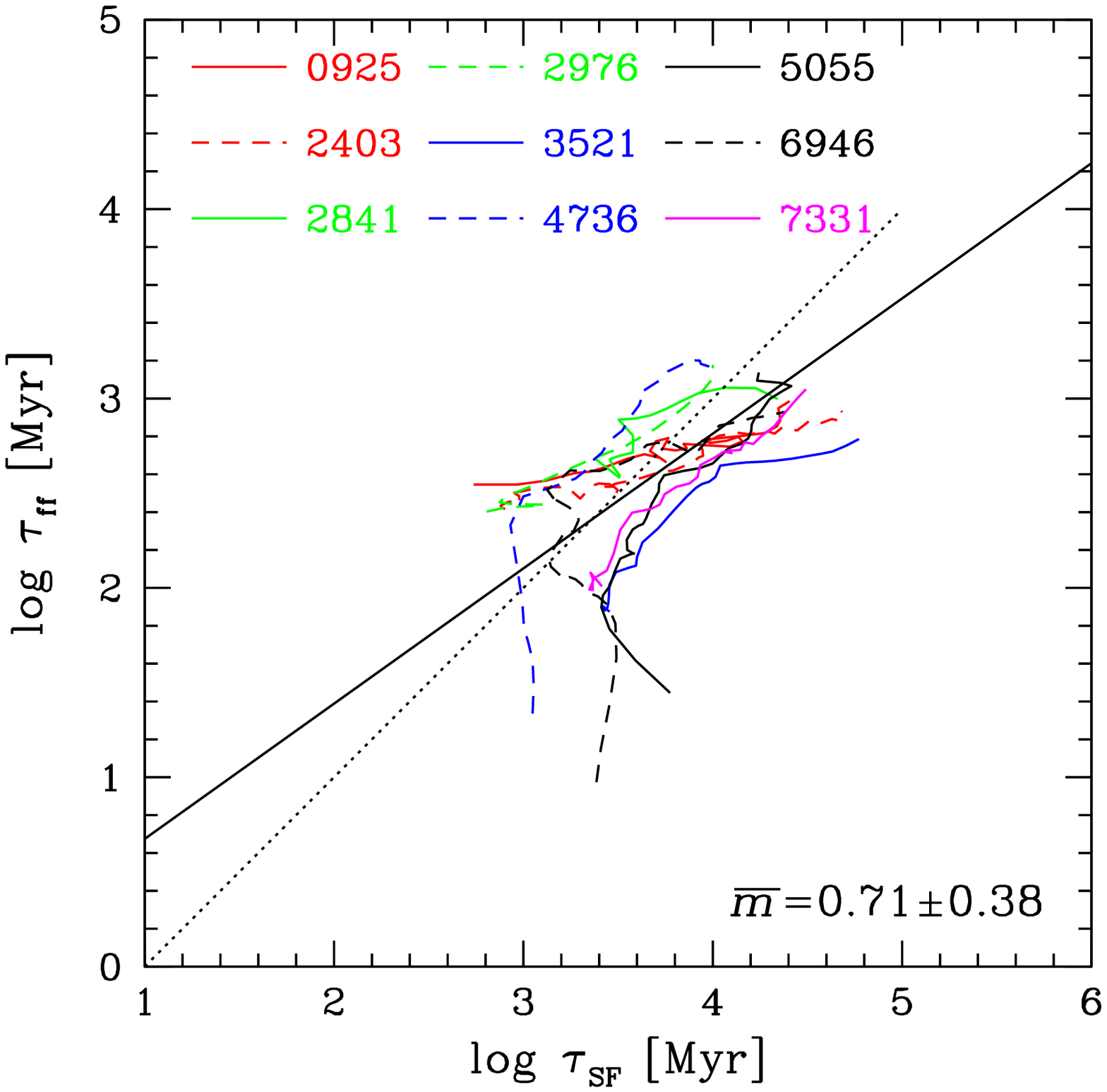}\hfill
\includegraphics[bb=36 156 540 665,clip,width={.32\textwidth}]{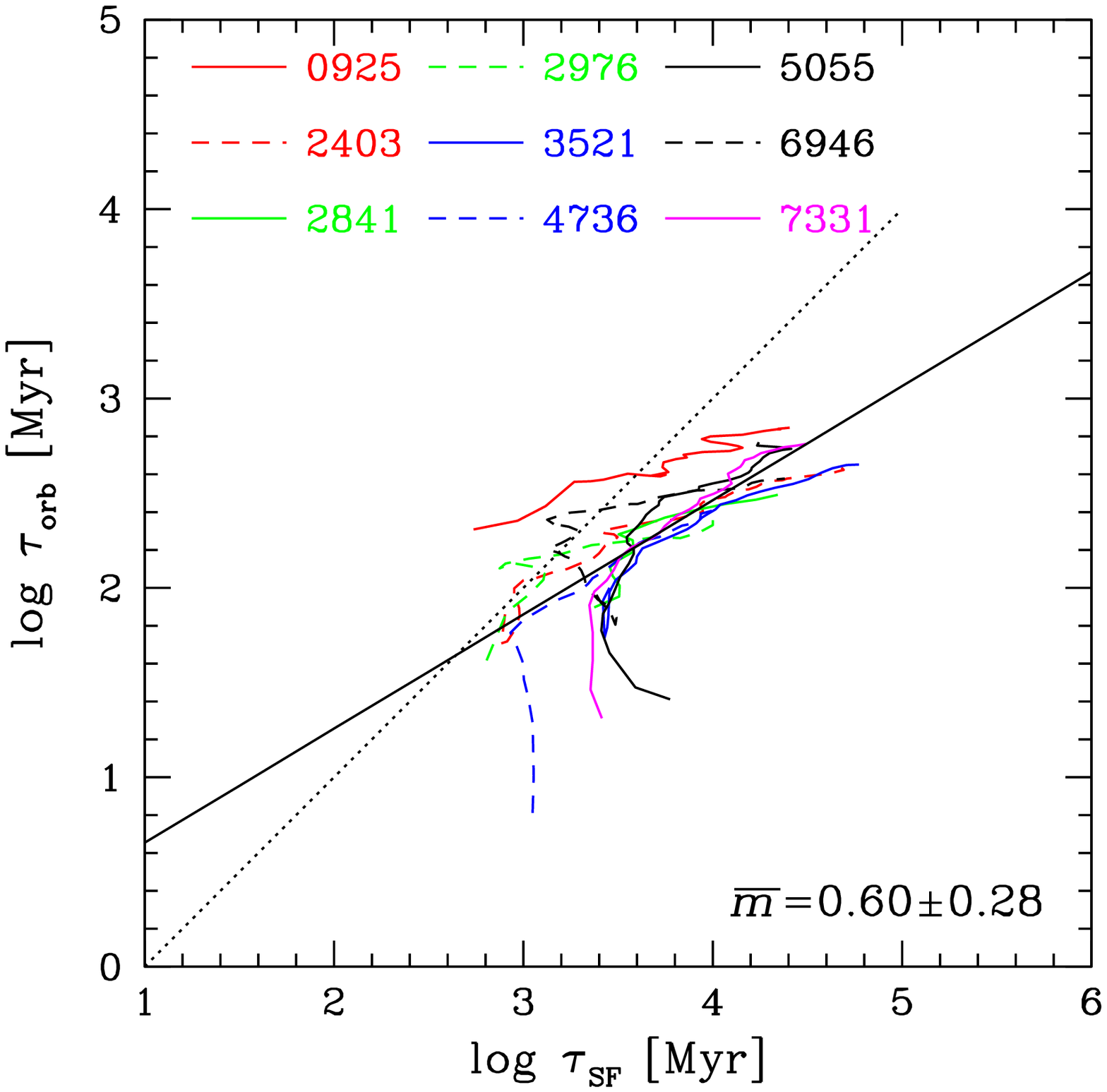}
\end{center}
\caption{Comparison of star formation timescale with (a) Jeans timescale for both stars and gas; (b) midplane free-fall timescale; (c) orbital timescale.  Data are from L08.  The solid line represents the average least-squares slope for the sample, while the dotted line represents a slope of unity.}
\label{fig:tcorplt}
\end{figure*}

\begin{figure*}
\begin{center}
\includegraphics[bb=36 156 540 665,clip,width={.32\textwidth}]{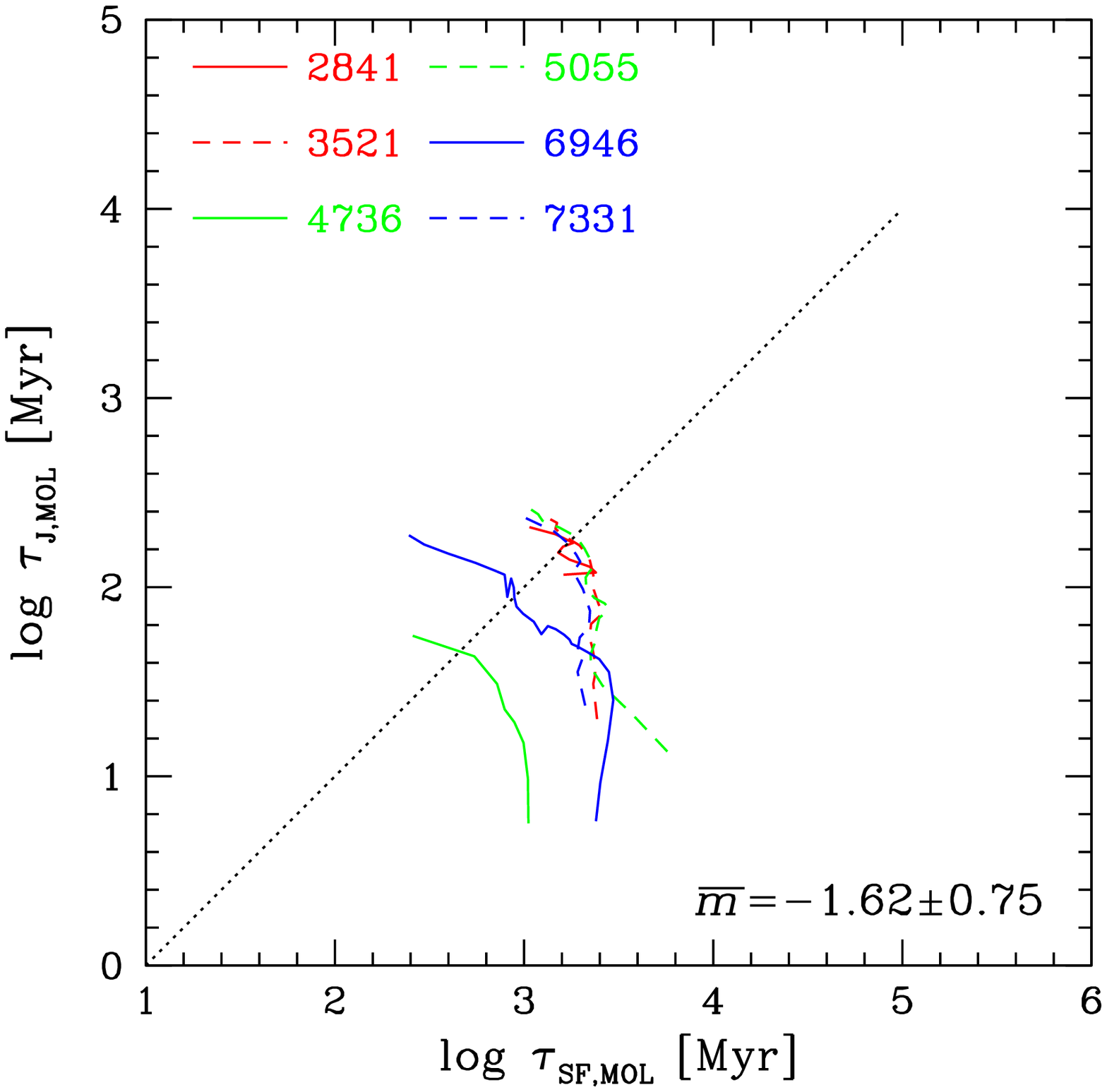}\hfill
\includegraphics[bb=36 156 540 665,clip,width={.32\textwidth}]{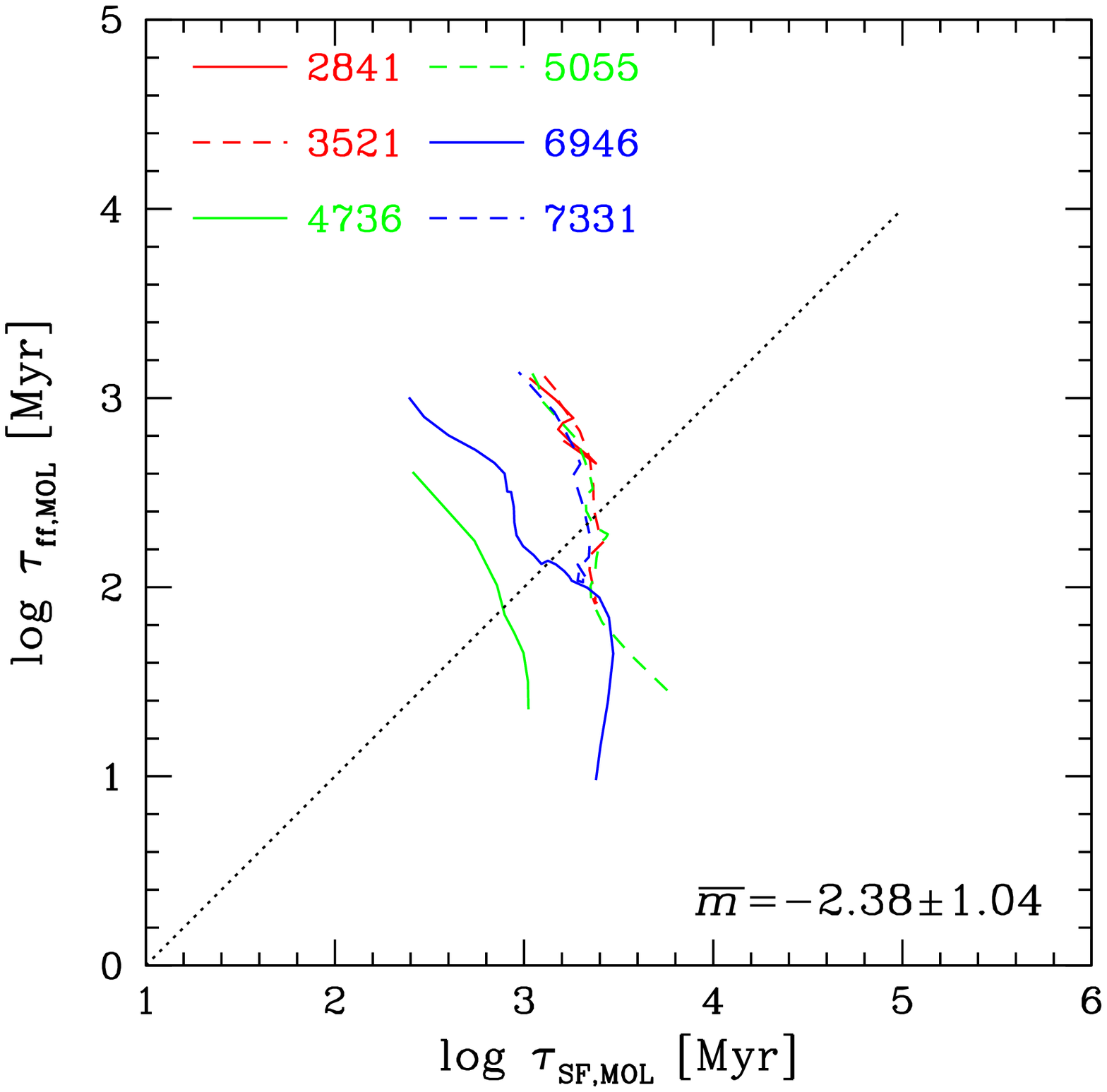}\hfill
\includegraphics[bb=36 156 540 665,clip,width={.32\textwidth}]{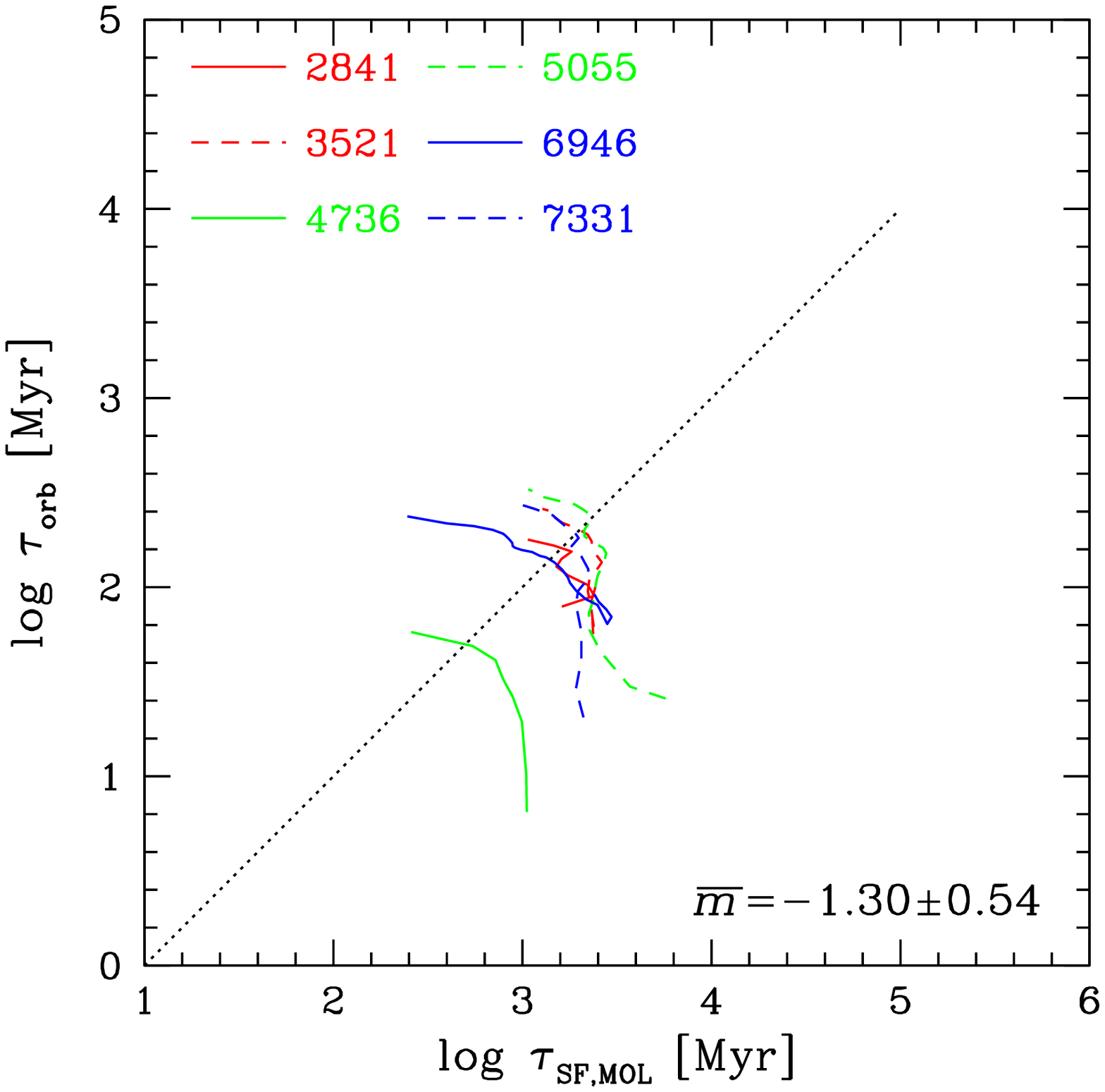}
\end{center}
\caption{Comparison of star formation timescale with (a) Jeans timescale for both stars and gas; (b) midplane free-fall timescale; (c) orbital timescale.  Data are from L08, and only the H$_2$ component of the gas is used to compute the timescales.  The dotted line represents a slope of unity.}
\label{fig:tcorplth2}
\end{figure*}

\begin{figure*}
\begin{center}
\includegraphics[bb=36 156 540 665,clip,width={.32\textwidth}]{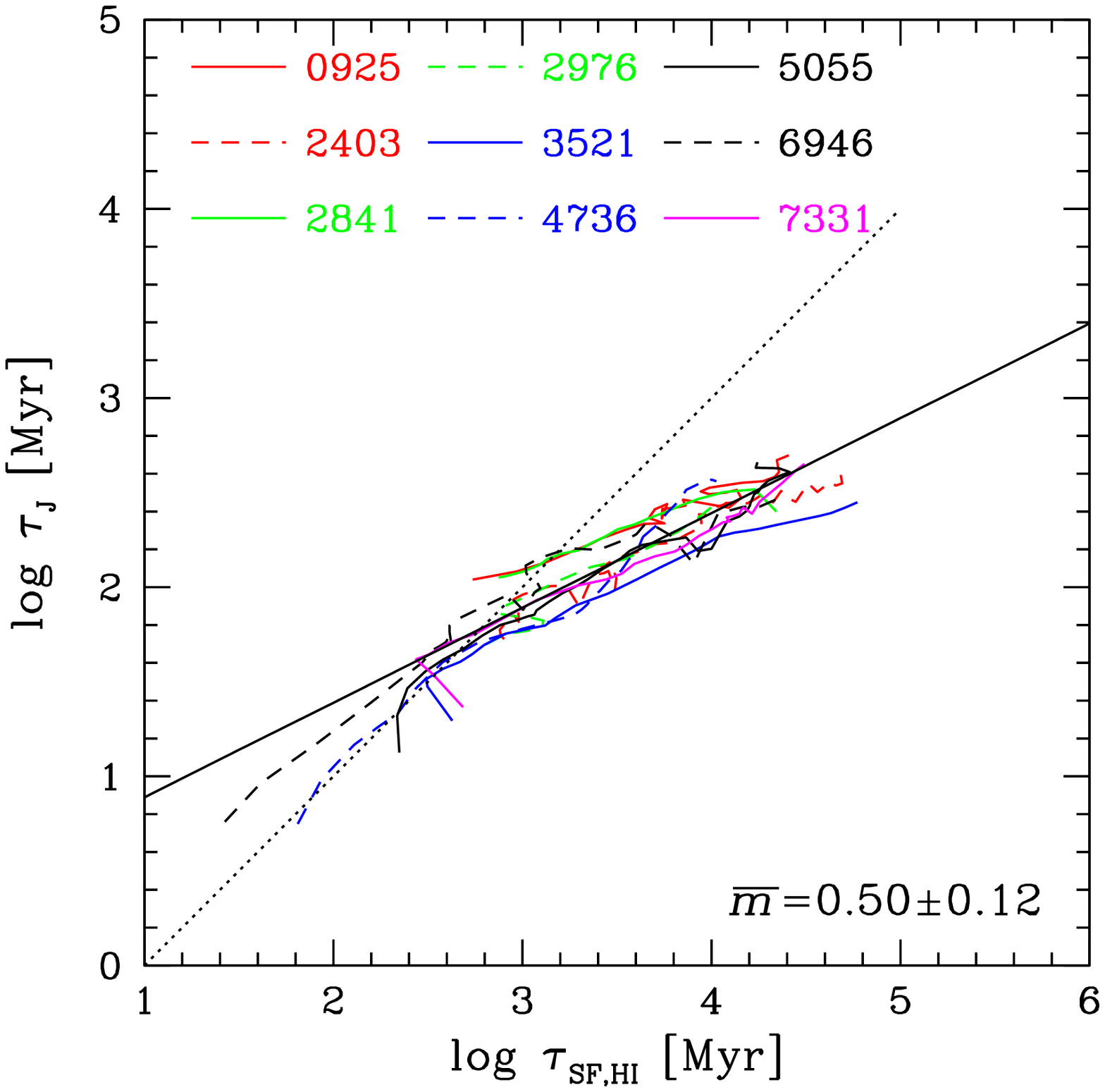}\hfill
\includegraphics[bb=36 156 540 665,clip,width={.32\textwidth}]{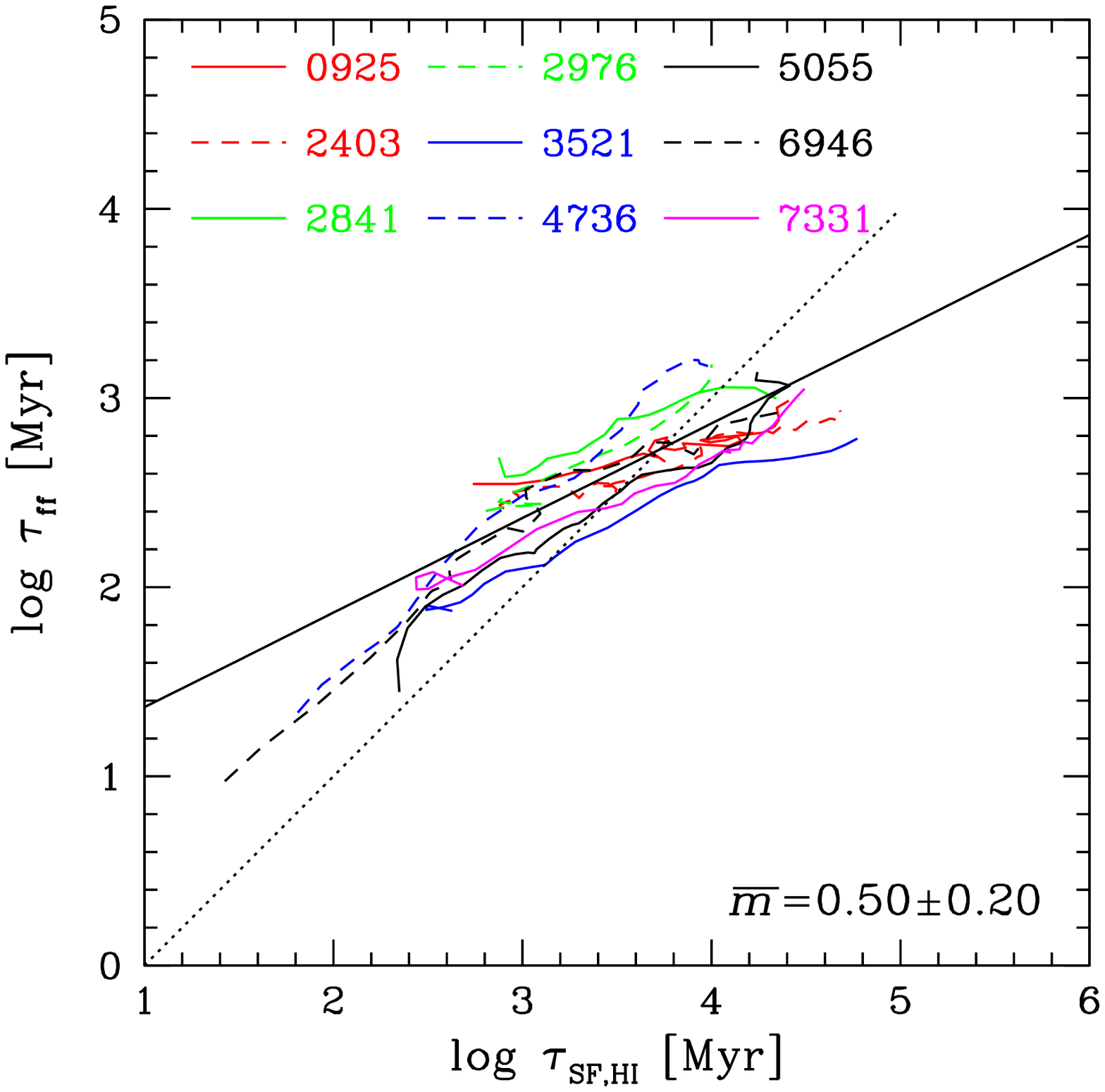}\hfill
\includegraphics[bb=36 156 540 665,clip,width={.32\textwidth}]{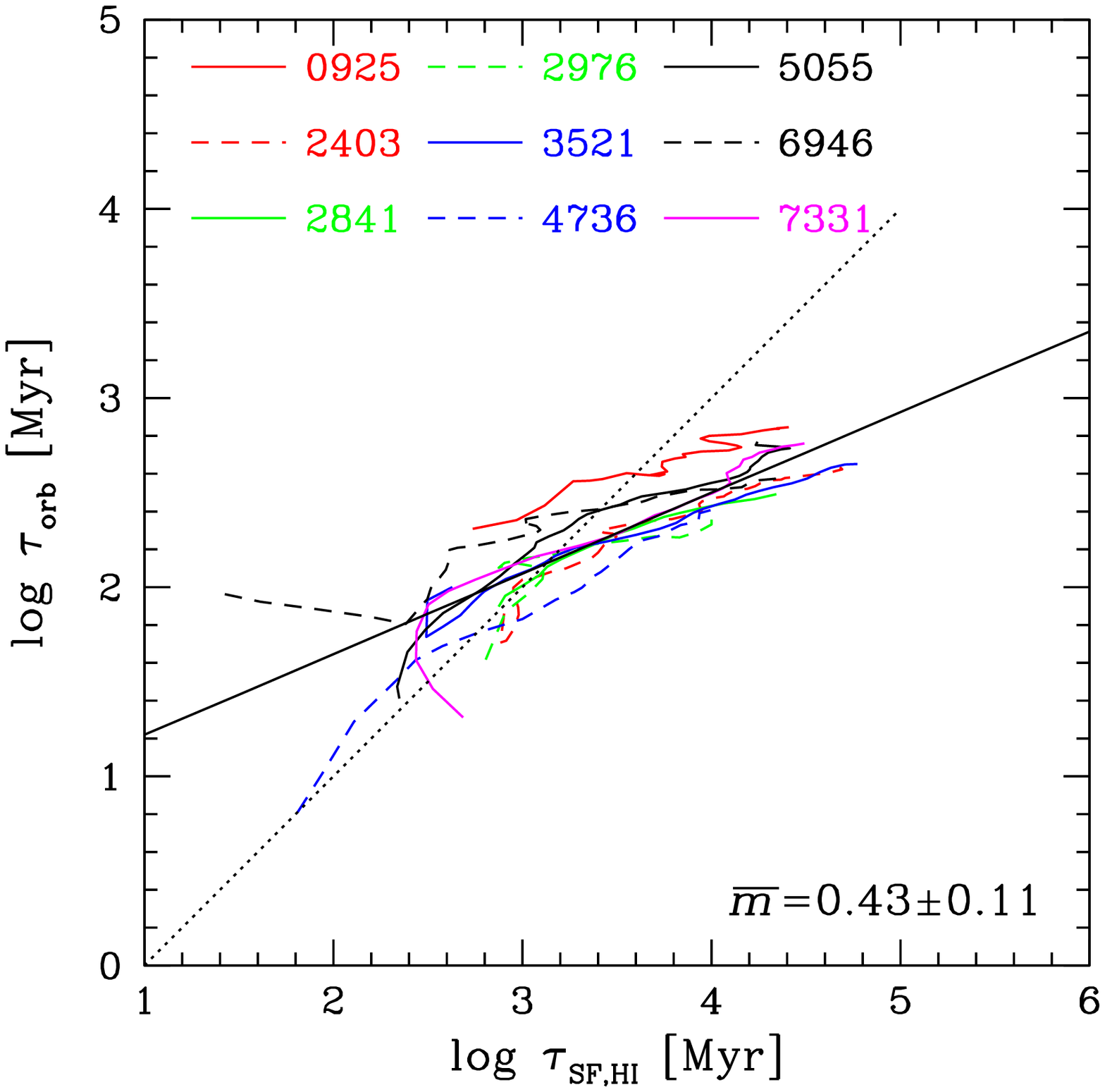}
\end{center}
\caption{Comparison of star formation timescale, using only the \HI\ component for the gas mass, with (a) Jeans timescale for both stars and gas; (b) midplane free-fall timescale; (c) orbital timescale.  Data are from L08.  The solid line represents the average least-squares slope for the sample, while the dotted line represents a slope of unity.}
\label{fig:tcorplthi}
\end{figure*}

\begin{table}
\begin{center}
\caption{Radial scalelengths for various timescales\label{tbl:rscale}}
\begin{tabular}{lcc}
\hline
& CO-faint sample & CO-bright sample\\
$l(\tau_{\rm SF})/R_{25}$ & 0.34 $\pm$ .08 & 0.48 $\pm$ .14\\
$l(\tau_{\rm J})/R_{25}$  & 0.65 $\pm$ .16 & 0.40 $\pm$ .06\\
$l(\tau_{\rm ff})/R_{25}$ & 0.79 $\pm$ .26 & 0.38 $\pm$ .09\\
$l(\tau_{\rm orb})/R_{25}$ & 0.76 $\pm$ .23 & 0.50 $\pm$ .13\\
\hline
\end{tabular}
\end{center}
\end{table}

\section{Results}\label{sec:results}

\subsection{Radial variation in timescales}

Figure~\ref{fig:tradplt}(a) shows the radial variation in the observed star formation timescale, $\tau_{\rm SF} \equiv \siggas/\sigsfr$, in comparison with the three dynamical timescales that we consider, based on the L08 data.  Timescales are derived from rates using the relation $\tau=2\pi/\omega$, with $\tau_{\rm orb}=2\pi/\Omega$.  Figure~\ref{fig:tradplt}(b) shows the same quantities but for the BIMA SONG data.  Different line styles denote different galaxies, whereas different colors denote different timescales.  Radii have been normalized by the optical radius $R_{25}$ as given by LEDA, typically $\approx$300\arcsec.  The CO-faint galaxies NGC 925, 2403, 2841, and 2976 are shown in the upper panels, while the CO-bright galaxies NGC 3521, 4736, 5055, 6946, and 7331 are shown in the lower panels.  Note that $\tau_{\rm SF}$ has been scaled by 0.1 to make it comparable to the other timescales.

Generally speaking, $\tau_{\rm SF}$ shows an increase with radius, from $\sim$1 Gyr in the central regions to $\sim$10 Gyr at the edge of the optical disk.  The central values of $\tau_{\rm SF}$ appear to be smaller in CO-faint galaxies; as discussed by L08, this provides indirect evidence that the H$_2$ content in these galaxies is underestimated, since it seems unlikely that CO-faint galaxies would have higher star formation efficiencies, given their low metallicities, strong radiation fields, and weak gravitational fields.  For the CO-bright galaxies, $\tau_{\rm SF}$ appears to flatten in the inner regions; in these regions the ISM is largely molecular, and a near proportionality between \sigsfr\ and \siggas\ is found (constant $\tau_{\rm SF}$).  NGC 6946 is a notable departure from these overall trends: for this galaxy $\tau_{\rm SF}$ reaches a minimum at intermediate radius of $\sim$0.5$R_{25}$, due to the very active star formation in its spiral arms.

Table~\ref{tbl:rscale} summarizes the radial exponential scalelengths for the various timescales, normalized to $R_{25}$.  These are obtained from unweighted least-squares fits to the curves in Figure~\ref{fig:tradplt}(a).  The most notable trend is that, for the CO-faint galaxies, the exponential scalelength for $\tau_{\rm SF}$ is a factor of $\sim$2 shorter than the scale lengths of all of the dynamical timescales.  Thus none of the dynamical timescales seems able to predict the star formation timescale in these galaxies.  For the CO-bright galaxies, on the other hand, the radial increase in $\tau_{\rm SF}$ is reasonably well matched by the radial increase in the dynamical timescales.

Figures~\ref{fig:jeans}(a) and \ref{fig:jeans}(b) show the effect of neglecting the stellar contribution to the Jeans rate when calculating $\tau_{\rm J}$.  Particularly in the CO-faint galaxies, where the radial gas profile is dominated by \HI\ with fairly constant surface density, considering only gas leads to a very flat profile for $\tau_{\rm J}$.  In CO-bright galaxies, the gas-only $\tau_{\rm J}$ profiles also show much more variation with radius, as expected given the considerable amount of substructure in the radial gas profiles.

Figures~\ref{fig:tnorm}(a) and \ref{fig:tnorm}(b) display the same data as in Fig.~\ref{fig:tradplt}, but now the ratio of $\tau_{\rm SF}$ to each of the dynamical timescales is plotted.  These plots reveal more clearly that in CO-faint galaxies, $\tau_{\rm SF}$ displays a steeper radial gradient than do any of the dynamical timescales.  For CO-bright galaxies, the ratios show little systematic variation with radius, except near small $R$ where the flattening of $\tau_{\rm SF}$ and the continued decline of $\tau_{\rm J}$ and $\tau_{\rm ff}$ is apparent.  

In general, the higher resolution SONG data [Figures~\ref{fig:tradplt}(b) and \ref{fig:tnorm}(b)] are consistent with the L08 profiles outside of the central regions ($R/R_{25} > 0.1$), as expected since the difference in resolution only becomes apparent on scales of $\lesssim$20\arcsec.  Closer to the center ($R/R_{25} < 0.1$), differences become apparent at high resolution.  As discussed in \S\ref{sec:behavior}, these trends may reflect limitations in estimating gas masses or dynamical timescales in these regions.  For the analysis in \S\ref{sec:corr}, we confine our analysis to the L08 data, which are more complete in radial coverage and are also well-matched in resolution.

\subsection{Correlations Between Timescales}\label{sec:corr}

Figure~\ref{fig:tcorplt} shows correlation plots between $\tau_{\rm SF}$ and each of the three dynamical timescales.  These plots demonstrate clearly the minimum value of $\tau_{\rm SF}$ (about 1--3 Gyr) achieved in the inner parts of the CO-bright galaxies (black, blue, and magenta curves).  In the CO-faint galaxies, the slope of the relation is generally flatter than unity, again reflecting the steeper radial gradient of $\tau_{\rm SF}$.  Because of the scaling factor of 0.1 applied to the abscissa, $\tau_{\rm SF}$ is always longer than any of the dynamical timescales considered.

Figure~\ref{fig:tcorplth2} shows similar correlation plots for the case in which only the H$_2$ gas is used to calculate the star formation, free-fall, and Jeans timescales.  Such a comparison is motivated by the possibility that \HI\ is inert as far as star formation is concerned, and that warm \HI\ in particular may not contribute to the gravitational instability of the disk.  We find no correlation between $\tau_{\rm SF,H_2}$ and any of the three dynamical timescales---indeed, they appear to be anti-correlated, particularly in NGC 4736 and 6946.  Thus, if star formation only involves H$_2$, there is no evidence that a dynamical timescale is relevant for establishing its rate.  The negative slopes in the correlation plots, which at face value suggest that molecular clouds produce stars with greater efficiency in the outer disk, may indicate that CO emission provides an incomplete inventory of H$_2$ mass in these regions.

Finally, in Figure~\ref{fig:tcorplthi} only the \HI\ component of \siggas\ is used to calculate the star formation timescale, though the H$_2$ is including in estimating the dynamical rates.  If star formation in giant molecular clouds (GMCs) occurs at a constant rate per unit H$_2$ mass, and if GMC lifetimes are not a function of environment, then any variation in the star formation timescale (within or among galaxies) should reflect the timescale for forming GMCs from \HI.  Thus in a steady-state situation, the rate at which \HI\ clouds form GMCs (e.g., the Jeans rate) should be proportional to the rate at which \HI\ would be consumed by the current SFR.  While all three dynamical timescales show a good correlation with $\tau_{\rm SF,HI}$, the correlations are much flatter than linear (power law slopes 0.4--0.5).  In other words, regions of low SFE possess dynamical times that are too short, and regions of high SFE possess dynamical times that are too long, compared to the simple steady-state model.  Of the three dynamical timescales, $\tau_{\rm J}$ appears to show the strongest correlation with $\tau_{\rm SF,HI}$.

\section{Discussion}

\subsection{Behavior of Timescales Near Galaxy Centers}\label{sec:behavior}

The higher angular resolution of the BIMA SONG data [Fig.~\ref{fig:tradplt}(b) and \ref{fig:tnorm}(b)] allows us to probe dynamical timescales closer to the centers of the galaxies, though we note that $\tau_{\rm SF}$ is still calculated using the lower resolution \sigsfr\ profile.  For several of the galaxies, such as NGC 4736, 5055, and 6946, both $\tau_{\rm J}$ and $\tau_{\rm ff}$ show a steep decline for $R$$<$30\arcsec\ ($R/R_{25}<0.1$) in the high-resolution data, reflecting the steep increase in $\Sigma_*$ associated with the stellar bulge.  While the bulge is an important additional term in the gravitational potential, we have almost certainly overestimated its effect by including it in $\Sigma_*$ but adopting a disk-based value for $c_*$.  A larger value of $c_*$ would serve to reduce $\omega_{\rm ff}$ and $\omega_{\rm J}$ in these regions and thus increase the corresponding timescales.  Moreover, the applicability of the Jeans time for a thin disk becomes increasingly suspect within the more spherical potential of a stellar bulge.  Thus, the short dynamical timescales in this region may be more indicative of limitations in our modeling than of conditions that are especially conducive to star formation.

The ratio of $\tau_{\rm SF}$ to the dynamical timescales shown in Fig.~\ref{fig:tnorm}(b) also shows the effect of the sharp central drop in $\tau_{\rm J}$ and $\tau_{\rm ff}$, but the ratio is enlarged even further by an apparent increase in $\tau_{\rm SF}$ near the centers of several galaxies (again, NGC 4736, 5055, and 6946 are prime examples).  Although this may be due in part to the mismatched resolution of the \sigsfr\ profiles, these trends are also exhibited to a lesser extent in the L08 data [Fig.~\ref{fig:tradplt}(a)].  A large reservoir of molecular gas with relatively little massive star formation has been noted in the central region of NGC 4736 \citep{Wong:00}, but other explanations exist as well, such as a change in the CO-to-H$_2$ conversion factor $X_{\rm CO}$ due to higher gas pressures (e.g., \citealt{Regan:01}; see also discussion in \S\ref{sec:variations} below).

Notable in Fig.~\ref{fig:tnorm}(b) is the sharp drop in $\tau_{\rm SF}$/$\tau_{\rm ff}$ in the central region of NGC 2841; this region exhibits very little neutral gas but a high rate of star formation, as seen in both FUV and 24$\mu$m emission (L08).  The dearth of gas led to the omission of this region from the radial profiles of L08, and hence it is also omitted from Figures~\ref{fig:tradplt}(a) and \ref{fig:tnorm}(a).  It is unclear whether this galaxy has experienced a recent starburst that has depleted the gas, or whether the central FUV and IR emission arises from a source other than star formation.

\subsection{Variations in the Ratio of Timescales}\label{sec:variations}

The results of \S\ref{sec:results} indicate that, ignoring any distinction between \HI\ and H$_2$, the star formation timescale cannot be modeled as simply proportional to a dynamical timescale.  In low-mass, CO-faint galaxies, the dynamical timescales fail to reproduce the steep radial gradient observed in $\tau_{\rm SF}$.  In CO-bright galaxies they fare somewhat better, but are unable to reproduce the constant $\tau_{\rm SF}$ found in the inner, H$_2$-dominated regions and also predict too shallow a radial gradient in the outer disk.  These conclusions were already reached by L08 for $\tau_{\rm ff}$ and $\tau_{\rm orb}$, so it is not surprising that they apply to $\tau_{\rm J}$ as well.

One possibility is that $\tau_{\rm SF}$ is systematically overestimated, due to an overestimate of the amount of star-forming gas or an underestimate of the SFR.  This would have to occur {\it both} in the inner, H$_2$-dominated disks of spirals (where the discrepancy is 0.6--0.8 dex) and in the outer disks of \HI-dominated dwarfs (where the discrepancy is 0.8--1 dex).  In the inner disks, a decrease in $X_{\rm CO}$ by a factor of 4--6, or an increase in the SFR by a similar factor, would help to linearize the relation between $\tau_{\rm J}$ and $\tau_{\rm SF}$.  There is some evidence for at least the first of these effects: a decrease in $X_{\rm CO}$ by factors of 2--5 has been inferred towards galaxy centers based on measurements of $^{13}$CO emission \citep{Paglione:01}.  On the other hand, it seems unlikely that SFRs are substantially underestimated by L08, since they employed a combination of FUV and IR data.  In outer disks, $\tau_{\rm SF}$ could be overestimated if a large fraction (84\%) of the \HI\ is inert as far as star formation is concerned (e.g., in the warm neutral phase), or if the stellar initial mass function (IMF) disfavors high-mass stars due to a paucity of high-mass clusters (\citealt{Weidner:05}, but see also \citealt{Elmegreen:06}).

Alternatively, the dynamical timescale $\tau_{\rm J}$ may be understimated, or equivalently $\omega_{\rm J}$ overestimated.  In outer disk regions, this could be symptomatic of a decline in the stellar mass-to-light ratio.  While \citet{Bell:03c} deduce a modest decrease in $M/L_K$ for bluer galaxy colors, the trend is highly dependent on metallicity, and is too small ($<$0.2 dex) to account for the observed values of $\tau_{\rm J}$.  

Another source of systematic error may be our crude approximations for the velocity dispersions $c_g$ and $c_*$.  It seems unlikely that an increase in $c_g$ or $c_*$ would occur in the outer disk (such trends have never been reported), but an unmodeled increase of one or both of these might lead us to underestimate $\tau_{\rm J}$ in inner, H$_2$-dominated regions.  As noted above (\S\ref{sec:behavior}), $c_*$ may increase faster than expected near galaxy centers due to the presence of a bulge.  An increase in the \HI\ velocity dispersion towards the centers of THINGS galaxies has been found by \citet{Tamburro:09}, though the increase appears too small (about a factor of 2) to offset the apparent decline in $\tau_{\rm J}$ as $R \rightarrow 0$.

In summary, while inclusion of a number of effects could act to linearize the relation between $\tau_{\rm SF}$ and $\tau_{\rm J}$, the most plausible appear to be a change in $X_{\rm CO}$ and $c_*$ in the inner disk and a change in the fraction of star-forming gas in the outer disk.  Investigating each of these possibilities remains an important, albeit difficult, observational challenge.

\subsection{Dual-Phase Approach}

The tighter correlation between $\tau_{\rm SF}$ when only \HI\ is included and $\tau_{\rm J}$ (Fig.~\ref{fig:tcorplthi}) suggests another possibility: that the dynamical timescale is important for the formation of H$_2$, but that other physical processes (presumably internal to molecular clouds) govern the timescale for star formation.  The simplest model of this type would be one in which GMC lifetimes are constant and the SFR per unit H$_2$ mass is also constant.  In that case the star formation rate would be proportional to the molecular cloud formation rate (CFR):
\begin{equation}
\sigsfr \propto \Sigma_{\rm CFR} \propto \frac {\Sigma({\rm HI})} {\tau_{\rm J}}\;.
\end{equation}
However, rearranging this implies that $\tau_{\rm SF,HI} \propto \tau_{\rm J}$, which is not observed.  

A slightly less restrictive model keeps the SFR per unit H$_2$ mass constant, and assumes a steady-state cycling between \HI\ and H$_2$:
\begin{equation}
\frac{\Sigma({\rm HI})}{\tau_{\rm HI \rightarrow H_2}} = \frac{\Sigma({\rm H_2})} {\tau_{\rm GMC, life}} \propto \frac{\sigsfr}{\tau_{\rm GMC, life}}\;.
\end{equation}
Under the assumption that $\tau_{\rm HI \rightarrow H_2} \propto \tau_{\rm J}$, this implies that
\begin{equation}
\tau_{\rm SF,HI} \equiv \frac{\Sigma({\rm HI})}{\sigsfr} \propto \frac{\tau_{\rm J}}{\tau_{\rm GMC, life}} \;.
\end{equation}
The (approximately) observed relation $\tau_{\rm J} \propto (\tau_{\rm SF,HI})^{0.5}$ would then require
\begin{equation}
\tau_{\rm GMC,life} \propto \tau_{\rm J}^{-1}\;.
\label{eqn:tauinv}
\end{equation}

Independent (since it does not rely on the SFR) support for the relation in Eq.~(\ref{eqn:tauinv}) comes from the observed empirical correlation
\begin{equation}
R_{\rm mol} \equiv \frac{\Sigma({\rm H_2})}{\Sigma({\rm HI})} \propto P_{\rm h} \;,
\label{eqn:ph}
\end{equation}
where the midplane hydrostatic pressure $P_{\rm h}$ is given by \citep{Elmegreen:89}:
\begin{equation}
P_{\rm h} \equiv \frac{\pi G\Sigma_g^2}{2}\left(1+\frac{c_g}{c_*}\frac{\Sigma_*}{\Sigma_g}\right) = \frac{c_g\siggas}{2\tau_{\rm J}} = \frac{c_g^2}{G\eta\tau_{\rm J}^2} \;.
\end{equation}
The proportionality between $R_{\rm mol}$ and $P_{\rm h}$, which holds approximately in the CO-bright galaxies examined by \citet{Wong:02} and \citet{Blitz:04}, implies that
\begin{equation}
R_{\rm mol} = \frac{\tau_{\rm GMC, life}}{\tau_{\rm HI \rightarrow H_2}} \propto \frac{1}{\tau_{\rm J}^2}\;,
\end{equation}
assuming that $c_g$ and $\eta$ do not vary significantly with radius.  Thus, if $\tau_{\rm HI \rightarrow H_2} \propto \tau_{\rm J}$, then $\tau_{\rm GMC,life} \propto \tau_{\rm J}^{-1}$.

Indeed, the tightness of the relation between $R_{\rm mol}$ and $P_{\rm h}$ is directly related to the tightness of the relations between $\tau_{\rm J}$ and $\tau_{\rm SF,HI}$ and between \sigsfr\ and $\Sigma({\rm H_2})$.  If $\sigsfr \propto \Sigma({\rm H_2})$ then it follows that $\tau_{\rm SF,HI} \propto R_{\rm mol}^{-1}$; at the same time, the definition of $P_h$ implies that $\tau_{\rm J}^2 \propto P_{\rm h}^{-1}$, setting aside the correction factor $\eta$.  This may provide a route for understanding the origin of the $R_{\rm mol} \propto P_{\rm h}$ relation in terms of molecular cloud formation and destruction, and for understanding the limitations to its wider applicability.  For instance, \citet{Koyama:09b} have found that in simulations where the $Q$ parameter is allowed to vary, the $R_{\rm mol} \propto P_{\rm h}$ relation no longer holds.  This suggests that galactic rotation has a strong influence on $\tau_{\rm HI \rightarrow H_2}$ or $\tau_{\rm GMC,life}$ which is obscured by the fact that $\omega_{\rm J} \approx \Omega$ in real galaxies.

Thus, one can account for the two empirical correlations, $\sigsfr \propto \Sigma({\rm H_2})$ and $\tau_{\rm J} \propto (\tau_{\rm SF,HI})^{0.5}$,
by assuming that \HI\ forms H$_2$ on a Jeans timescale, and that the reverse process of dissociating H$_2$ occurs on a timescale that scales {\it inversely} with the Jeans timescale.  In principle this could be a viable scenario, since conditions that favor rapid molecular cloud formation (such as high external pressure, or a high gas shielding column) could also forestall cloud destruction.  For instance, if the expansion of \HII\ regions is related to the destruction of molecular clouds, then since the stall radius scales inversely with external pressure $P$, cloud lifetimes may therefore scale with $P$.  However, such a scaling would be difficult to reconcile with the expectation that high-mass clouds are more quickly unbound by the massive stars which form within them \citep{Matzner:02}.  Further work will be needed to cast Eq.~(\ref{eqn:tauinv}) in a dimensionally correct form and to derive the observed proportionality constant.

\begin{figure}
\includegraphics[bb=30 165 580 680,clip,width={.48\textwidth}]{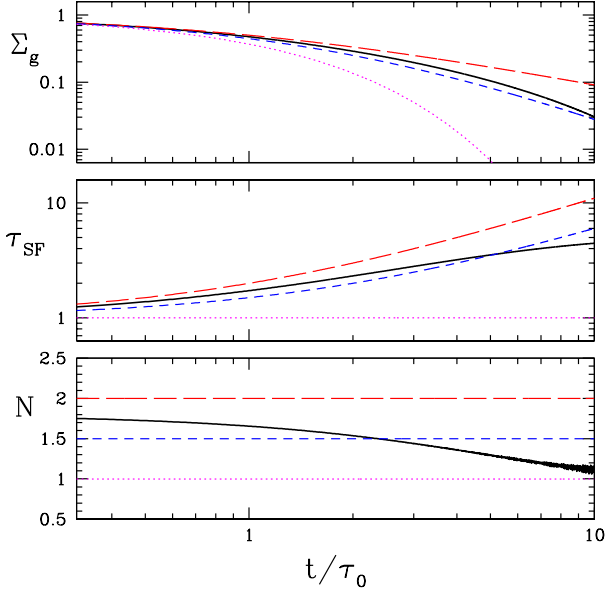}
\caption{Time evolution of \siggas, $\tau_{\rm SF}$, and the instantaneous Schmidt law index $N=d\ln\Sigma_{\rm SFR}/d\ln\Sigma_g$ for a closed-box model (solid line) where the star formation timescale is proportional to the Jeans time, and $c_g/c_*$ is fixed at 0.2.  The gas surface density and star formation timescale are normalized to their initial values, and $\tau_0$ is the initial star formation timescale.  Constant Schmidt law indices of 1 ({\it magenta dotted line}), 1.5 ({\it blue short-dashed line}), and 2 ({\it red long-dashed line}) are shown for comparison.}
\label{fig:model}
\end{figure}

\subsection{Comparison with the Schmidt Law}\label{sec:schmidt}

Although we have not established a linear relationship between the observed star formation timescale and the modified Jeans timescale $\tau_{\rm J}$, it is nonetheless interesting to consider the implications such a relationship would have on the observed star formation law.
For simplicity we consider a closed-box model, where all the gas is in place at $t$=0.  By assumption the total mass surface density at a given radius is constant: $\Sigma(r,t) = \Sigma_0(r)$.  Defining the gas fraction as $f_g(r,t) = \siggas(r,t)/\Sigma_0(r)$, simple integration of the classical Schmidt law given by Eq.~(\ref{eqn:kslaw}) yields
$$
f_g = \left\{ 
    \begin{array}{cl}
    \left[1+kt(N-1)(\Sigma_0)^{N-1}\right]^{-1/(N-1)} & \quad N\ne 1 \\
    \exp(-kt) & \quad N=1
    \end{array} \right.
$$
\citep[cf.][]{Bell:00}.  In terms of the local star formation timescale,
\[\tau_0 = \frac{\Sigma_0}{\Sigma_{\rm SFR,0}} = \frac{1}{k(\Sigma_0)^{N-1}}\;,\]
this can be written as:
\begin{equation}
f_g = \left\{ 
    \begin{array}{cl}
    \left[1+(N-1)t/\tau_0\right]^{-1/(N-1)} & \quad N\ne 1 \\
    \exp(-t/\tau_0) & \quad N=1
    \end{array} \right.
\end{equation}

If instead we adopt a star formation law based on the modified Jeans rate $\omega_{\rm J}$, such that
\begin{equation}
\sigsfr \propto \frac{\Sigma_g^2}{c_g}\left(1+\frac{c_g}{c_*}\frac{\Sigma_*}{\Sigma_g}\right)\;,
\end{equation}
then once again starting from a pure gas disk, 
\begin{equation}
\frac{df_g}{dt} = -\frac{f_g^2}{\tau_0}\left(1+\frac{c_g}{c_*}\frac{1-f_g}{f_g}\right)\;.
\end{equation}
We have integrated this equation numerically for $\tau_0$=1 Gyr and a fixed ratio $c_g/c_*$=0.2.  Figure~\ref{fig:model} shows the time evolution of the gas and star formation surface densities, normalized to their initial values.  Note that the evolution of \siggas\ is quite similar to what would be expected for an $N$=1.5 Schmidt law.  However, the measured Schmidt law index $N$ varies with time, from an initial value close to 2 (appropriate for a pure gas disk) to a value close to 1 at late times (when the gas is almost entirely consumed).  This simple example illustrates that the observed Schmidt law index may be a function of the current evolutionary state of a galaxy. 

\subsection{Relevance of Timescales Estimated from Average Densities}

The average gas densities used to derive $\tau_{\rm J}$ and $\tau_{\rm ff}$ in this paper are often much less than the surface and volume densities of giant molecular clouds ($\sim$100 M$_\odot$ pc$^{-2}$ and $\sim$100 cm$^{-3}$ respectively) where the bulk of star formation actually occurs.  One may certainly question whether such average densities can be reliably used to infer the SFR.  We adopt this as a working hypothesis, assuming that either (1) large-scale average densities correlate well with local GMC densities, as would be the case if both scale with ambient pressure; or (2) the star formation rate is ultimately limited by the GMC formation rate, which reflects the conditions of the ISM on much larger scales.  The fact that $Q_{\rm eff} \approx 1$ in observed disks offers indirect evidence that large-scale dynamics influences star formation, since it is presumably star formation feedback that increases $c_g$ and thus $Q$ to values that are marginally unstable.  Unfortunately, the GMC formation rate is difficult to infer directly, since the ages of clouds cannot be easily determined, and the overall HI-H$_2$ balance depends on both H$_2$ formation and destruction rates.  Detailed correlation studies of young stellar objects and molecular clouds in the nearest galaxies, such as the Large Magellanic Cloud, should be able to better constrain these parameters.

Nonetheless, we cannot exclude the possibility that the star formation and dynamical timescales may not scale linearly because they are determined over regions with very different densities.  Recently, \citet{Koyama:09a} have conducted numerical simulations of galactic disks including stellar feedback, and compared the large-scale dynamical timescales ($\tau_{\rm ff}$, $\tau_{\rm orb}$, and $\tau_{\rm J,gas}$) with star formation timescales derived by assuming a constant star formation efficiency above a density threshold.  They also find a non-linearity between $\tau_{\rm SF}$ and $\tau_{\rm ff}$, which they attribute to estimating the dynamical timescales on spatial scales much larger than star-forming clouds.  In particular, $\tau_{\rm SF}/\tau_{\rm ff}$ increases in regions of high \siggas, similar to what we observe in H$_2$-dominated regions.  Intuitively this makes sense, since $\tau_{\rm SF}$ becomes constant as the threshold density is approached, whereas $\tau_{\rm ff}$ continues to decrease.  On the other hand, it is not clear whether such simulations can account for the minimum in the $\tau_{\rm SF}/\tau_{\rm ff}$ ratio at intermediate radii and its subsequent increase in the outer, \HI-dominated disk.

Finally, we note that any theory relating dynamical timescales with $\tau_{\rm SF}$ must explain the order of magnitude difference between them.  The difficulty in accounting for this low efficiency factor has been discussed extensively in the literature \citep[e.g.,][]{Elmegreen:02}.  One possibility is that this factor arises from a universal density structure generated by turbulence, coupled with a threshold density for star formation that is a fixed multiple of the mean density.  This effectively ensures that once H$_2$ forms, only a small, roughly constant fraction of it participates in star formation.  Variations of this idea have been explored by \citet{Elmegreen:02} and \citet{Krumholz:05}, but the appropriate value for the threshold density has yet to be clearly established.

\section{Conclusions}

We have determined the dynamical timescales in nearby disk galaxies using three common estimators, modified to account for the gravity of the stellar disk.  The Jeans and orbital timescales are related by an effective $Q$ parameter, which is observed to be close to 1 in real galaxies, making the two timescales interchangeable.  Use of the modified Jeans rate to derive the SFR leads to a star formation law that resembles the $N$=1.5 Schmidt law, but which, for a closed-box model, has a value of $N$ which declines over time.  On the other hand, recent observational data suggest that the star formation rate is not simply proportional to the modified Jeans rate, although inclusion of the stellar component does improve the correlation.  Several effects, including a change in the $X_{\rm CO}$ factor in the inner disk, improper modeling of the stellar velocity dispersion of the bulge, and a decrease in the fraction of star-forming gas in the outer disk, could contribute to the non-linearity.  Future studies using the far-infrared dust emission to derive $X_{\rm CO}$ and resolving the cloudy structure of the neutral ISM may be able to quantify these effects.

Alternatively, we consider a two-step model where the Jeans rate determines the GMC formation rate, whereas GMCs form stars at a constant rate.  In agreement with this model, we find a much tighter correlation between the \HI\ depletion timescale (due to star formation) and the Jeans timescale, but again it is not linear; to linearize it would require that the GMC (or H$_2$) destruction rate is {\it inversely correlated} with the Jeans rate.  Although somewhat unexpected, such a scaling would also account for the correlation between the H$_2$/\HI\ ratio and the midplane hydrostatic pressure.  Observations of GMCs and their embedded stellar populations in Local Group galaxies will be needed to investigate this possibility.

\acknowledgments

We thank Erwin de Blok for providing the THINGS rotation curves in tabular format, and Adam Leroy for helpful feedback.  This research was supported by the National Science Foundation through grant AST-0838226 to the Combined Array for Research in Millimeter Astronomy (CARMA) and by the University of Illinois.

\bibliographystyle{apj}
\bibliography{merged}

\begin{thebibliography}{43}
\expandafter\ifx\csname natexlab\endcsname\relax\def\natexlab#1{#1}\fi

\bibitem[{{Banerjee} \& {Jog}(2007)}]{Banerjee:07}
{Banerjee}, A. \& {Jog}, C.~J. 2007, \apj, 662, 335

\bibitem[{Bell \& {de Jong}(2000)}]{Bell:00}
Bell, E.~F. \& {de Jong}, R.~S. 2000, MNRAS, 312, 497

\bibitem[{{Bell} {et~al.}(2003){Bell}, {McIntosh}, {Katz}, \&
  {Weinberg}}]{Bell:03c}
{Bell}, E.~F., {McIntosh}, D.~H., {Katz}, N., \& {Weinberg}, M.~D. 2003, \apjs,
  149, 289

\bibitem[{{Bigiel} {et~al.}(2008){Bigiel}, {Leroy}, {Walter}, {Brinks}, {de
  Blok}, {Madore}, \& {Thornley}}]{Bigiel:08}
{Bigiel}, F., {Leroy}, A., {Walter}, F., {Brinks}, E., {de Blok}, W.~J.~G.,
  {Madore}, B., \& {Thornley}, M.~D. 2008, \aj, 136, 2846

\bibitem[{{Blitz} \& {Rosolowsky}(2004)}]{Blitz:04}
{Blitz}, L. \& {Rosolowsky}, E. 2004, \apjl, 612, L29

\bibitem[{Boissier {et~al.}(2003)Boissier, Prantzos, Boselli, \&
  Gavazzi}]{Boissier:03}
Boissier, S., Prantzos, N., Boselli, A., \& Gavazzi, G. 2003, MNRAS, 346, 1215

\bibitem[{{de Blok} {et~al.}(2008){de Blok}, {Walter}, {Brinks},
  {Trachternach}, {Oh}, \& {Kennicutt}}]{deBlok:08}
{de Blok}, W.~J.~G., {Walter}, F., {Brinks}, E., {Trachternach}, C., {Oh},
  S.-H., \& {Kennicutt}, R.~C. 2008, \aj, 136, 2648

\bibitem[{{Elmegreen}(1989)}]{Elmegreen:89}
{Elmegreen}, B.~G. 1989, \apj, 338, 178

\bibitem[{Elmegreen(1992)}]{Elm:92b}
Elmegreen, B.~G. 1992, in Star Formation, Galaxies, and the Interstellar
  Medium, ed. J.~Franco, F.~Ferrini, \& G.~Tenorio-Tagle (Cambridge: Cambridge
  U. Press), 337

\bibitem[{{Elmegreen}(2002)}]{Elmegreen:02}
---. 2002, \apj, 577, 206

\bibitem[{{Elmegreen}(2006)}]{Elmegreen:06}
---. 2006, \apj, 648, 572

\bibitem[{{Gao} \& {Solomon}(2004)}]{Gao:04b}
{Gao}, Y. \& {Solomon}, P.~M. 2004, \apj, 606, 271

\bibitem[{{Gil de Paz} {et~al.}(2007){Gil de Paz}, {Boissier}, {Madore},
  {Seibert}, {Joe}, {Boselli}, {Wyder}, {Thilker}, {Bianchi}, {Rey}, {Rich},
  {Barlow}, {Conrow}, {Forster}, {Friedman}, {Martin}, {Morrissey}, {Neff},
  {Schiminovich}, {Small}, {Donas}, {Heckman}, {Lee}, {Milliard}, {Szalay}, \&
  {Yi}}]{GildePaz:07}
{Gil de Paz}, A., {Boissier}, S., {Madore}, B.~F., {Seibert}, M., {Joe}, Y.~H.,
  {Boselli}, A., {Wyder}, T.~K., {Thilker}, D., {Bianchi}, L., {Rey}, S.-C.,
  {Rich}, R.~M., {Barlow}, T.~A., {Conrow}, T., {Forster}, K., {Friedman},
  P.~G., {Martin}, D.~C., {Morrissey}, P., {Neff}, S.~G., {Schiminovich}, D.,
  {Small}, T., {Donas}, J., {Heckman}, T.~M., {Lee}, Y.-W., {Milliard}, B.,
  {Szalay}, A.~S., \& {Yi}, S. 2007, \apjs, 173, 185

\bibitem[{{Helfer} {et~al.}(2003){Helfer}, {Thornley}, {Regan}, {Wong},
  {Sheth}, {Vogel}, {Blitz}, \& {Bock}}]{Helfer:03}
{Helfer}, T.~T., {Thornley}, M.~D., {Regan}, M.~W., {Wong}, T., {Sheth}, K.,
  {Vogel}, S.~N., {Blitz}, L., \& {Bock}, D.~C.-J. 2003, \apjs, 145, 259

\bibitem[{{Heyer} {et~al.}(2004){Heyer}, {Corbelli}, {Schneider}, \&
  {Young}}]{Heyer:04}
{Heyer}, M.~H., {Corbelli}, E., {Schneider}, S.~E., \& {Young}, J.~S. 2004,
  \apj, 602, 723

\bibitem[{{Jarrett} {et~al.}(2003){Jarrett}, {Chester}, {Cutri}, {Schneider},
  \& {Huchra}}]{Jarrett:03}
{Jarrett}, T.~H., {Chester}, T., {Cutri}, R., {Schneider}, S.~E., \& {Huchra},
  J.~P. 2003, \aj, 125, 525

\bibitem[{Kennicutt(1989)}]{KC:89}
Kennicutt, R.~C. 1989, ApJ, 344, 685

\bibitem[{Kennicutt(1998)}]{KC:98a}
---. 1998, ApJ, 498, 541

\bibitem[{{Kennicutt} {et~al.}(2003)}]{Kennicutt:03}
{Kennicutt}, R.~C. {et~al.} 2003, \pasp, 115, 928

\bibitem[{{Kennicutt} {et~al.}(2007){Kennicutt}, {Calzetti}, {Walter}, {Helou},
  {Hollenbach}, {Armus}, {Bendo}, {Dale}, {Draine}, {Engelbracht}, {Gordon},
  {Prescott}, {Regan}, {Thornley}, {Bot}, {Brinks}, {de Blok}, {de Mello},
  {Meyer}, {Moustakas}, {Murphy}, {Sheth}, \& {Smith}}]{Kennicutt:07}
{Kennicutt}, Jr., R.~C., {Calzetti}, D., {Walter}, F., {Helou}, G.,
  {Hollenbach}, D.~J., {Armus}, L., {Bendo}, G., {Dale}, D.~A., {Draine},
  B.~T., {Engelbracht}, C.~W., {Gordon}, K.~D., {Prescott}, M.~K.~M., {Regan},
  M.~W., {Thornley}, M.~D., {Bot}, C., {Brinks}, E., {de Blok}, E., {de Mello},
  D., {Meyer}, M., {Moustakas}, J., {Murphy}, E.~J., {Sheth}, K., \& {Smith},
  J.~D.~T. 2007, \apj, 671, 333

\bibitem[{{Koyama} \& {Ostriker}(2009{\natexlab{a}})}]{Koyama:09a}
{Koyama}, H. \& {Ostriker}, E.~C. 2009{\natexlab{a}}, \apj, 693, 1316

\bibitem[{{Koyama} \& {Ostriker}(2009{\natexlab{b}})}]{Koyama:09b}
---. 2009{\natexlab{b}}, \apj, 693, 1346

\bibitem[{{Kregel} {et~al.}(2002){Kregel}, {van der Kruit}, \& {de
  Grijs}}]{Kregel:02}
{Kregel}, M., {van der Kruit}, P.~C., \& {de Grijs}, R. 2002, \mnras, 334, 646

\bibitem[{{Krumholz} \& {McKee}(2005)}]{Krumholz:05}
{Krumholz}, M.~R. \& {McKee}, C.~F. 2005, \apj, 630, 250

\bibitem[{{Krumholz} \& {Thompson}(2007)}]{Krumholz:07}
{Krumholz}, M.~R. \& {Thompson}, T.~A. 2007, \apj, 669, 289

\bibitem[{{Leroy} {et~al.}(2008){Leroy}, {Walter}, {Brinks}, {Bigiel}, {de
  Blok}, {Madore}, \& {Thornley}}]{Leroy:08}
{Leroy}, A.~K., {Walter}, F., {Brinks}, E., {Bigiel}, F., {de Blok}, W.~J.~G.,
  {Madore}, B., \& {Thornley}, M.~D. 2008, \aj, 136, 2782

\bibitem[{{Matzner}(2002)}]{Matzner:02}
{Matzner}, C.~D. 2002, \apj, 566, 302

\bibitem[{{Olling}(1996)}]{Olling:96}
{Olling}, R.~P. 1996, \aj, 112, 457

\bibitem[{{Paglione} {et~al.}(2001){Paglione}, {Wall}, {Young}, {Heyer},
  {Richard}, {Goldstein}, {Kaufman}, {Nantais}, \& {Perry}}]{Paglione:01}
{Paglione}, T.~A.~D., {Wall}, W.~F., {Young}, J.~S., {Heyer}, M.~H., {Richard},
  M., {Goldstein}, M., {Kaufman}, Z., {Nantais}, J., \& {Perry}, G. 2001,
  \apjs, 135, 183

\bibitem[{Regan {et~al.}(2001)Regan, Thornley, Helfer, Sheth, Wong, Vogel,
  Blitz, \& Bock}]{Regan:01}
Regan, M.~W., Thornley, M.~D., Helfer, T.~T., Sheth, K., Wong, T., Vogel,
  S.~N., Blitz, L., \& Bock, D. C.-J. 2001, ApJ, 561, 218

\bibitem[{Schmidt(1959)}]{Schmidt:59}
Schmidt, M. 1959, ApJ, 129, 243

\bibitem[{{Shostak} \& {van der Kruit}(1984)}]{Shostak:84}
{Shostak}, G.~S. \& {van der Kruit}, P.~C. 1984, \aap, 132, 20

\bibitem[{{Talbot} \& {Arnett}(1975)}]{Talbot:75}
{Talbot}, Jr., R.~J. \& {Arnett}, W.~D. 1975, \apj, 197, 551

\bibitem[{{Tamburro} {et~al.}(2009){Tamburro}, {Rix}, {Leroy}, {Low}, {Walter},
  {Kennicutt}, {Brinks}, \& {de Blok}}]{Tamburro:09}
{Tamburro}, D., {Rix}, H.-W., {Leroy}, A.~K., {Low}, M.-M.~M., {Walter}, F.,
  {Kennicutt}, R.~C., {Brinks}, E., \& {de Blok}, W.~J.~G. 2009, \aj, 137, 4424

\bibitem[{{van der Kruit}(1988)}]{vdK:88}
{van der Kruit}, P.~C. 1988, \aap, 192, 117

\bibitem[{{van der Kruit} \& {Searle}(1981)}]{vdK:81a}
{van der Kruit}, P.~C. \& {Searle}, L. 1981, \aap, 95, 105

\bibitem[{{Walter} {et~al.}(2008){Walter}, {Brinks}, {de Blok}, {Bigiel},
  {Kennicutt}, {Thornley}, \& {Leroy}}]{Walter:08}
{Walter}, F., {Brinks}, E., {de Blok}, W.~J.~G., {Bigiel}, F., {Kennicutt},
  R.~C., {Thornley}, M.~D., \& {Leroy}, A. 2008, \aj, 136, 2563

\bibitem[{Wang \& Silk(1994)}]{Wang:94}
Wang, B. \& Silk, J. 1994, ApJ, 427, 759

\bibitem[{{Weidner} \& {Kroupa}(2005)}]{Weidner:05}
{Weidner}, C. \& {Kroupa}, P. 2005, \apj, 625, 754

\bibitem[{Wong \& Blitz(2000)}]{Wong:00}
Wong, T. \& Blitz, L. 2000, ApJ, 540, 771

\bibitem[{Wong \& Blitz(2002)}]{Wong:02}
---. 2002, ApJ, 569, 157

\bibitem[{{Wu} {et~al.}(2005){Wu}, {Evans}, {Gao}, {Solomon}, {Shirley}, \&
  {Vanden Bout}}]{Wu:05}
{Wu}, J., {Evans}, II, N.~J., {Gao}, Y., {Solomon}, P.~M., {Shirley}, Y.~L., \&
  {Vanden Bout}, P.~A. 2005, \apjl, 635, L173

\bibitem[{{Zuckerman} \& {Evans}(1974)}]{Zuckerman:74}
{Zuckerman}, B. \& {Evans}, II, N.~J. 1974, \apjl, 192, L149

\end{thebibliography}

\end{document}